\documentclass[preprint,12pt]{emulateapj}
\usepackage{natbib}
\usepackage[colorlinks=true,urlcolor=blue,citecolor=blue,linkcolor=blue,breaklinks]{hyperref}
\begin{document}

\newcommand{\nnhp}{N$_2$H$^+$}
\newcommand{\nthp}{N$_2$H$^+$}
\newcommand{\kms}{km$\,$s$^{-1}$}

\title{Interferometric Observations of High-Mass Star-Forming Clumps with Unusual N$_2$H$^+$/HCO$^+$ Line Ratios}
\author{Ian W. Stephens\altaffilmark{1}, James M. Jackson\altaffilmark{1},
Patricio Sanhueza\altaffilmark{1,2}, J. Scott Whitaker\altaffilmark{3}, Sadia Hoq\altaffilmark{1}, Jill M. Rathborne\altaffilmark{4},
Jonathan B. Foster\altaffilmark{1,5}}
\altaffiltext{1}{\itshape Institute for Astrophysical Research, Boston University, Boston, MA 02215, USA
ianws@bu.edu}
\altaffiltext{2}{\itshape Now at National Astronomical Observatory of Japan, 2-21-1 Osawa, Mitaka, Tokyo 181-8588, Japan}
\altaffiltext{3}{\itshape Physics Department, Boston University, Boston, MA 02215, USA}
\altaffiltext{4}{\itshape Australia Telescope National Facility, CSIRO Astronomy and Space Science, Epping, NSW, Australia}
\altaffiltext{5}{\itshape Yale Center for Astronomy and Astrophysics, New Haven, CT 06520, USA}

\interfootnotelinepenalty=10000

\begin{abstract}
The Millimetre Astronomy Legacy Team 90 GHz (MALT90) survey has detected high-mass star-forming clumps with anomalous N$_2$H$^+$/HCO$^+$(1--0) integrated intensity ratios that are either unusually high (``N$_2$H$^+$ rich") or unusually low (``N$_2$H$^+$ poor"). With 3 mm observations from the Australia Telescope Compact Array (ATCA), we imaged two N$_2$H$^+$ rich clumps, G333.234--00.061 and G345.144--00.216, and two N$_2$H$^+$ poor clumps, G351.409+00.567 and G353.229+00.672. In these clumps, the N$_2$H$^+$ rich anomalies arise from extreme self-absorption of the HCO$^+$ line.  G333.234--00.061 contains two of the most massive protostellar cores known with diameters of less than 0.1~pc, separated by a projected distance of only 0.12~pc. Unexpectedly, the higher mass core appears to be at an earlier evolutionary stage than the lower mass core, which may suggest that two different epochs of high-mass star formation can occur in close proximity. Through careful analysis of the ATCA observations and MALT90 clumps (including the G333, NGC~6334, and NGC 6357 star formation regions), we find that N$_2$H$^+$ poor anomalies arise at clump-scales and are caused by lower relative abundances of N$_2$H$^+$ due to the distinct chemistry of \ion{H}{2} regions or photodissociation regions.
\end{abstract}
\subjectheadings{stars: formation -- stars: massive -- astrochemistry -- ISM: clouds -- ISM: molecules}

\maketitle

\section{Introduction} Ê\label{sec:intro}
The formation of high-mass stars ($>$8~M$_{\sun}$) is not as well understood as the formation of low-mass stars \citep[e.g.,][]{Zinnecker2007}. Part of the reason is that high-mass stars are more disruptive to their natal environment due to large accretion rates, strong winds, and ionizing photons. The chemistry of star-forming regions mostly depends on the temperature and density, but in high-mass star-forming regions the effects of shocks and photochemistry become important as well. High-mass star-forming regions are rarer and more deeply embedded compared with their low-mass counterparts, further adding to the difficulty in understanding the formation processes which alter the chemistry of the natal environment of high-mass stars.

High-mass stars form within high-mass cores (defined here to have a size-scale of $\lesssim0.1$~pc and mass $<$~100~$M_\sun$) that arise in high-mass clumps (defined here to be $\sim$1~pc and $\gtrsim200$~$M_\sun$). Since high-mass stars form in regions with large visual extinctions, observations at wavelengths longer than visible light are needed to determine their locations and study their formation processes. Continuum surveys of the Galactic plane have been undertaken at multiple wavelengths, including GLIMPSE \citep[$Spitzer$ IRAC 3.6, 4.5, 5.8, and 8.0~$\mu$m,][]{Benjamin2003}, MIPSGAL \citep[$Spitzer$ MIPS 24 and 70~$\mu$m,][]{Carey2009}, Hi-GAL \citep[$Herschel$ PACS/SPIRE 70, 160, 250, 350, and 500~$\mu$m,][]{Molinari2010}, and ATLASGAL \citep[APEX, 870~$\mu$m,][]{Schuller2009}. The ATLASGAL survey was particularly successful at identifying high-mass star-forming clumps at a variety of evolutionary stages \citep{Foster2011}. 

Based on the locations of the ATLASGAL clumps, the Millimetre Astronomy Legacy Team 90 GHz survey \citep[MALT90,][]{Foster2011,Jackson2013,Foster2013}  used Mopra to observe 16 line transitions to explore the chemical composition and kinematic information of $\sim$3000 of these clumps. The catalog for MALT90 sources is given in J. Rathborne et al. (in preparation), which provides velocities, integrated intensities, and linewidths of each line for every observed ATLASGAL clump. Moreover, J. Rathborne et al. (in preparation) classifies each source based on its infrared emission (using GLIMPSE and MIPSGAL) as Quiescent, Protostellar, \ion{H}{2} region, Photodissociation Region (PDR), or Unknown (details are in Section \ref{sec:classsourceselection}).

Using the first year of MALT90 data, \citet{Hoq2013} explored the chemical evolution of clumps and found several ``N$_2$H$^+$ anomalies". These sources showed unexpectedly extreme integrated intensity ratios between N$_2$H$^+$(1--0) and HCO$^+$(1--0). ``N$_2$H$^+$ rich" sources were defined to have integrated intensity ratios [I(N$_2$H$^+$)/I(HCO$^+$)] greater than 4 and ``N$_2$H$^+$ poor" sources have [I(N$_2$H$^+$)/I(HCO$^+$)] less than 0.3. 

Chemical models for star-forming regions suggest that extreme values of the integrated intensity ratio [I(N$_2$H$^+$)/I(HCO$^+$)] can occur as the star-forming region evolves. During the prestellar phase, low temperatures coupled with high densities cause CO to freeze onto dust grains, while nitrogen-bearing molecules remain relatively unaffected \citep[e.g.,][]{Charnley1997,Bergin1997}. As star formation progresses, the temperature rises, and CO is released from the dust grains and destroys N$_2$H$^+$ molecules through the chemical reaction N$_2$H$^+$~+~CO~$\rightarrow$~HCO$^+$~+~N$_2$ \citep[e.g.,][]{Busquet2011}. Thus it may be expected that cold, dense sources are N$_2$H$^+$ rich, while warmer \ion{H}{2} regions and PDRs are N$_2$H$^+$ poor.

To investigate these intensity anomalies, we used the Australia Telescope Compact Array (ATCA) to observe two N$_2$H$^+$ rich sources and two N$_2$H$^+$ poor sources. We discuss the classification of MALT90 sources, the ATCA source selection, and the data reduction in Section \ref{sec:datareduc}. We analyze the N$_2$H$^+$ rich sources in Section \ref{sec:n2hprich} and find that G333.234--00.061 contains two very high-mass protostellar cores. We discuss the two observed N$_2$H$^+$ poor sources and develop a sample of all MALT90 N$_2$H$^+$ poor sources in Section \ref{sec:n2hppoor}. Finally, we summarize the findings in Section \ref{sec:summary}.

\section{Classification, Source Selection, Observations, and Data Reduction} \label{sec:datareduc}
\subsection{MALT90 Classification of Sources} \label{sec:classification}
J. Rathborne et al. (in preparation) classified all observed MALT90 sources as Quiescent, Protostellar, \ion{H}{2} region, PDR, or Unknown. This classification was done by eye and was based on two different $Spitzer$ mid-infrared RGB images: (1) 3.6, 4.5, and 8.0~$\mu$m (GLIMPSE), and (2) 3.6, 8.0, and 24~$\mu$m (GLIMPSE+MIPSGAL). The classification is based solely on the emission contained within the Mopra MALT90 beam (38$\arcsec$) centered at the location of an ATLASGAL source. An example of the classifying scheme can be seen in Figure 1 of \citet{Jackson2013}. The ATCA and MOPRA dishes have the same diameter and thus the ATCA primary beam is also well described by the MALT90 classification.

``Quiescent" (or prestellar) sources have no obvious emission at 3.6, 4.5, 8.0, and 24~$\mu$m and appear as dark extinction features at these mid-IR wavelengths. ``Protostellar" clumps show a compact region of 24~$\mu$m emission usually associated with more extended regions of enhanced 4.5 $\mu$m emission \citep[``green fuzzies,"][]{Chambers2009}. ``\ion{H}{2} regions" contain high-mass stars with strong UV radiation, causing emission to be bright and extended in all bands. They are particularly bright in the 8~$\mu$m (due to fluorescent excitation of polycyclic aromatic hydrocarbons) and 24 $\mu$m bands, causing these sources to appear yellow in the GLIMPSE+MIPSGAL images. ``PDR" sources are at the interface of molecular and ionized gas and show extended emission in all bands. They are especially bright in the 8~$\mu$m band and consist of well-defined PDR ridges. Finally, sources that did not fit into these four main categories were classified as ``Unknown". These clumps usually had no obvious feature in the $Spitzer$ images or contained a combination of extinction features intermixed with patchy 8~$\mu$m emission.

\citet{Hoq2013} found that for the year 1 MALT90 data, dust temperatures derived from $Herschel$/Hi-GAL continuum data correlate with this clump classification scheme, i.e., the temperature rises from Quiescent to Protostellar to \ion{H}{2}/PDR. This correlation provides confidence for the clump evolutionary stage assignment. 

\subsection{Selection and Parameters of Anomalous Sources}\label{sec:classsourceselection}
To investigate the N$_2$H$^+$ anomalies, we chose two N$_2$H$^+$ rich sources, AGAL333.234--00.061 and AGAL345.144--00.216, and two N$_2$H$^+$ poor sources, AGAL351.409+00.567 and AGAL353.229+00.672. For simplicity in the rest of this paper, we refer to these sources by their ``G" names  (G333.234--00.061, G345.144--00.216, G351.409+00.567, and G353.229+00.672) rather than their ATLASGAL (AGAL) names.

\begin{figure*}[ht!]
\plottwo{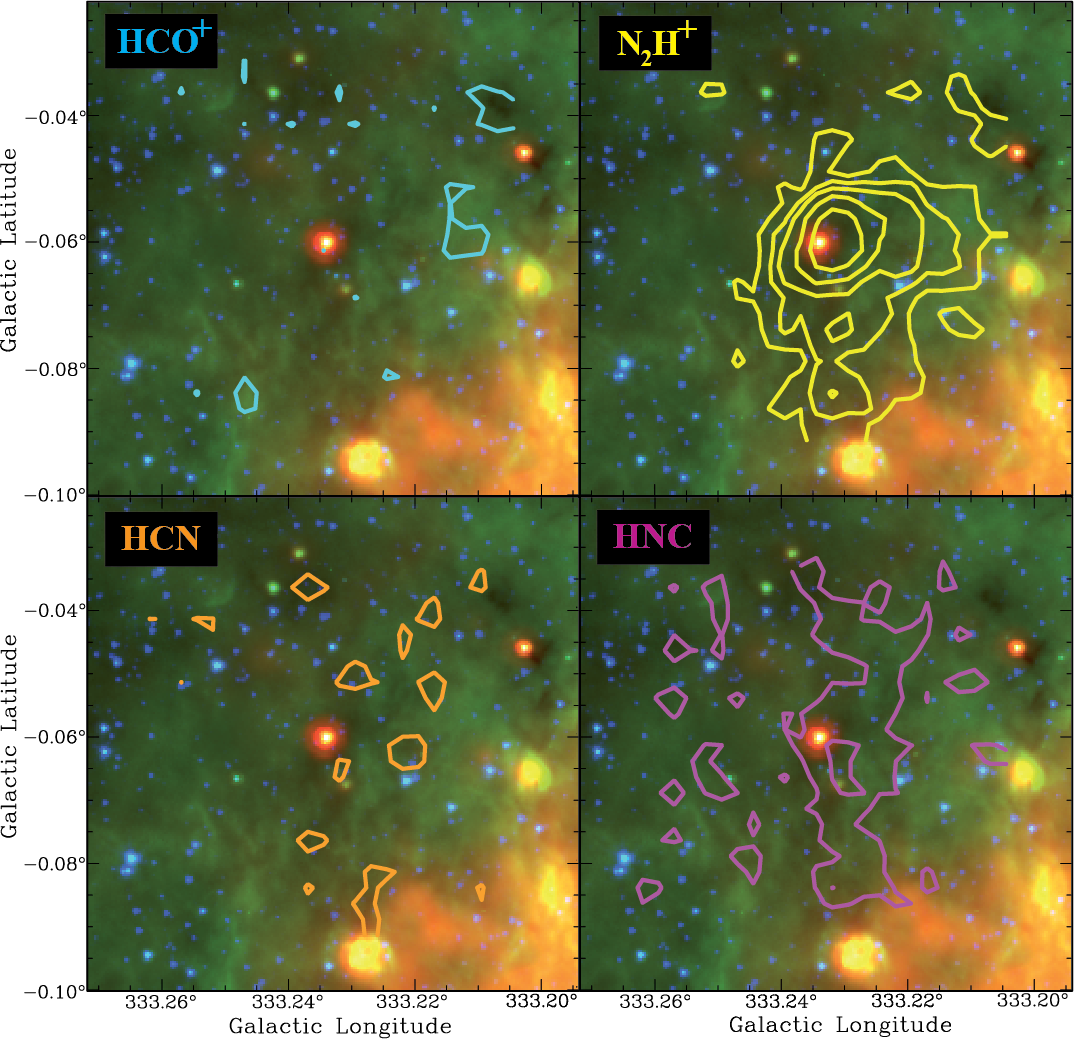}{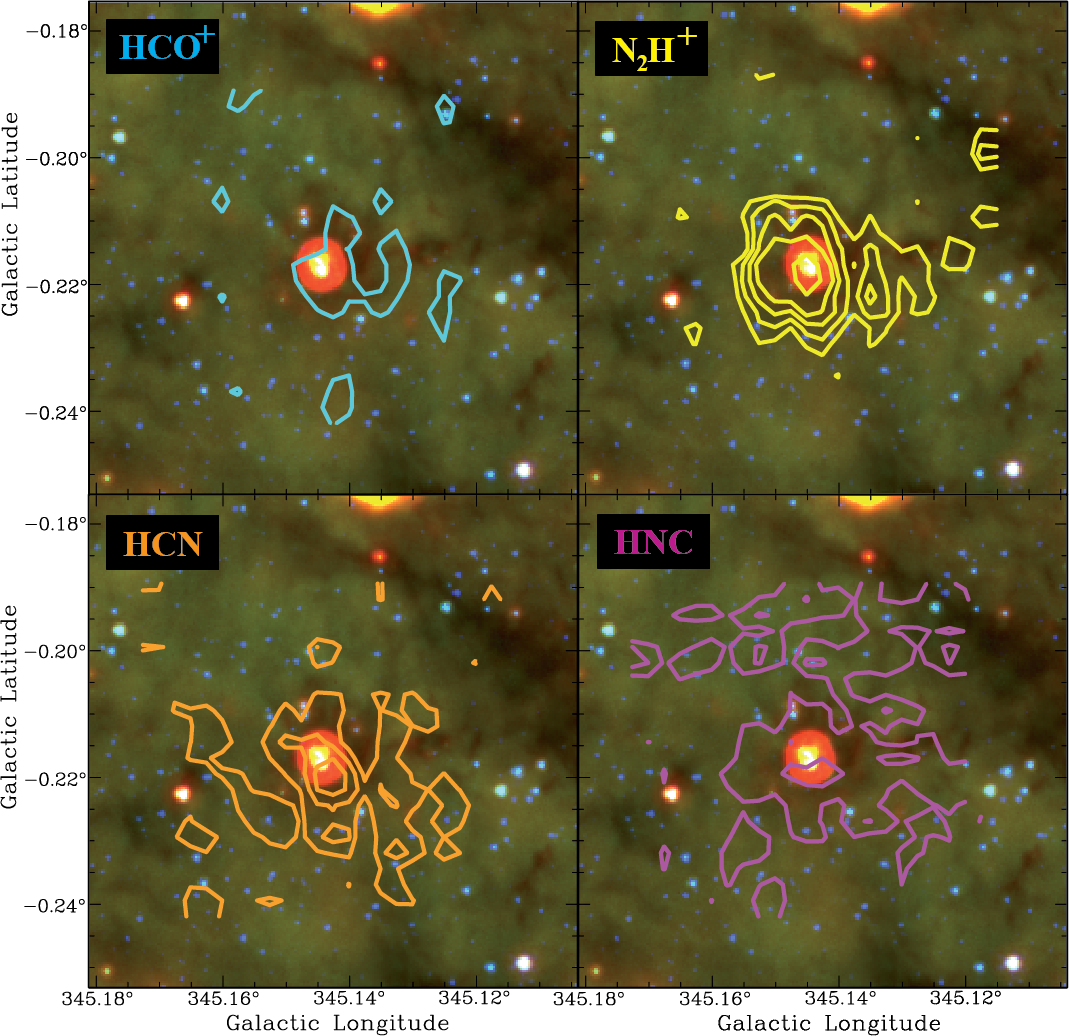}
\caption{$Spitzer$ three-color image (3.6~$\mu$m (blue), 8.0~$\mu$m (green), and 24~$\mu$m(red)) of the two N$_2$H$^+$ rich sources overlaid with MALT90 HCO$^+$(1--0), N$_2$H$^+$(1--0), HCN(1--0), and HNC(1--0) contours. Left: G333.234--00.061 with contours shown for [3, 6, 9, 15, and 24]~$\times$~$\sigma$, where $\sigma$ = 0.36~K~km$\,$s$^{-1}$. Right: G345.144--00.216 with contours shown for [3, 5, 7, 9, and 13]~$\times$~$\sigma$, where $\sigma$ = 0.37~K~km$\,$s$^{-1}$. \label{richanomalies}
}
\end{figure*}

\begin{figure*}[ht!]
\plottwo{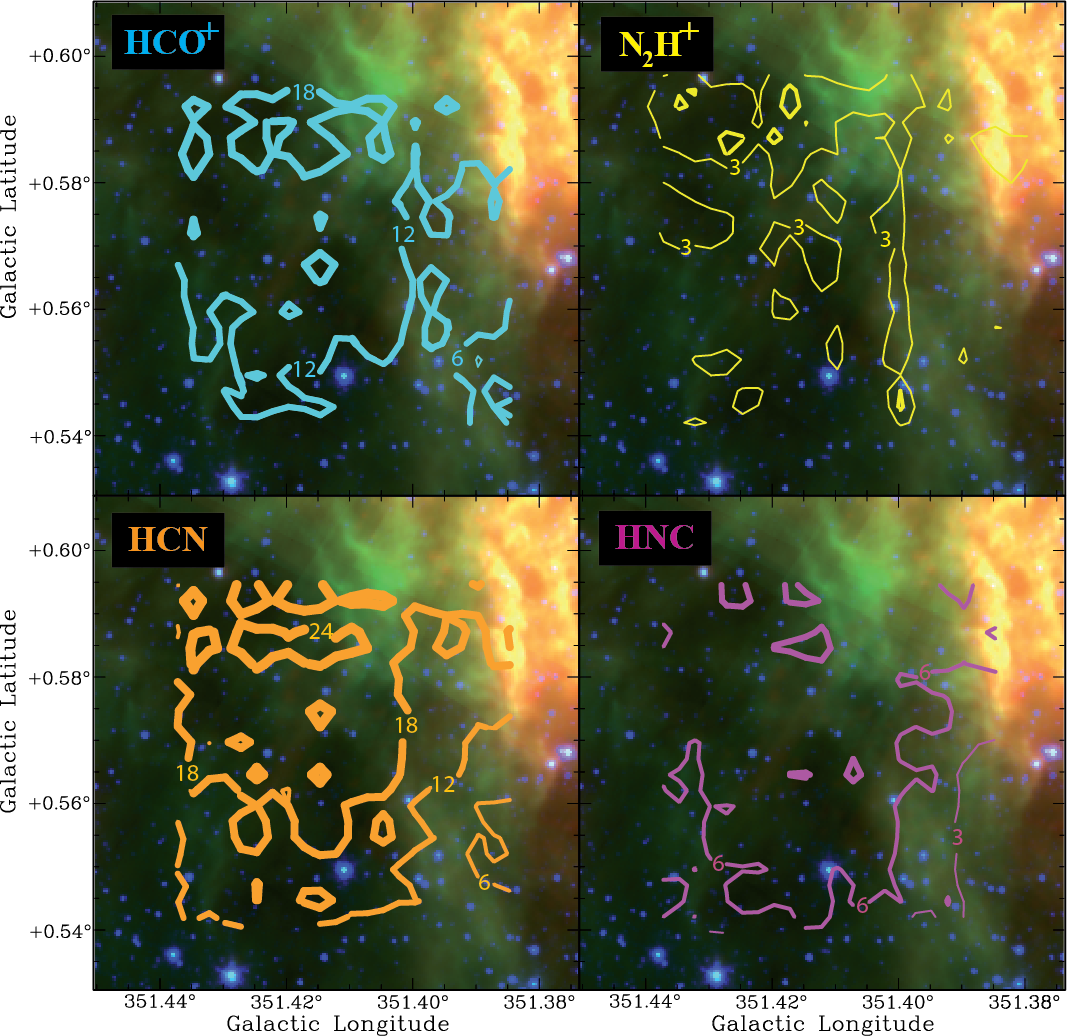}{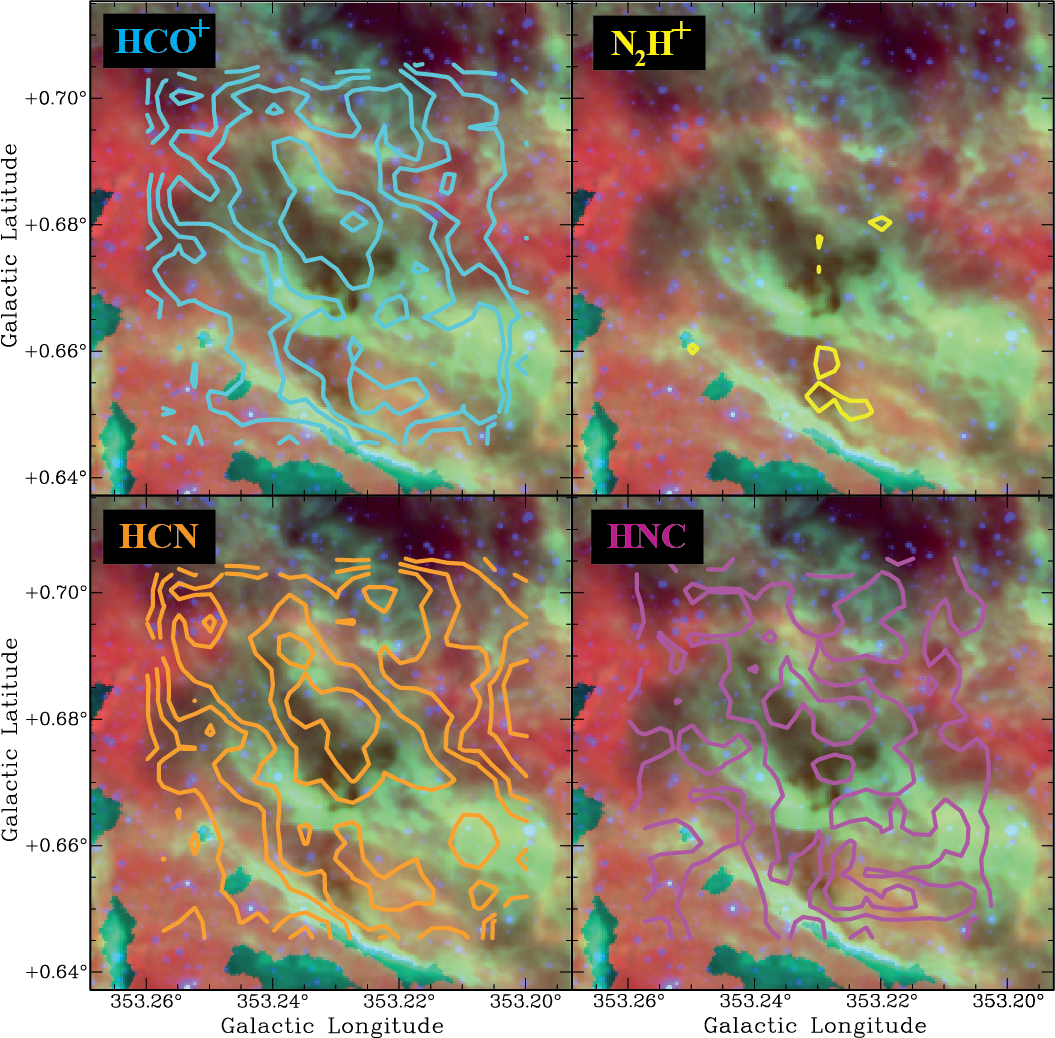}
\caption{$Spitzer$ three-color images (3.6~$\mu$m (blue), 8.0~$\mu$m (green), and 24~$\mu$m (red)) of the two N$_2$H$^+$ poor sources overlaid with Mopra MALT90 HCO$^+$(1--0), N$_2$H$^+$(1--0), HCN(1--0), and HNC(1--0) contours.  Left: G351.409+00.567 with contours shown for [3, 6, 12, 18, and 24]~$\times$~$\sigma$, where $\sigma$ = 0.49~K~km$\,$s$^{-1}$. For clarity, some contours are numbered based on their signal-to-noise, and thicker contours indicate more statistically significant contours. Note that HCO$^+$ and HCN show significantly stronger emission than \nthp. Right: G353.229+00.672 with contours shown for [3, 6, 9, 12, 18, and 24]~$\times$~$\sigma$, where $\sigma$ = 0.33~K~km$\,$s$^{-1}$. The green artifacts on the bottom left of this image are due to saturation in the MIPS 24~$\mu$m band. \label{pooranomalies} 
}
\end{figure*}

\begin{figure*}[ht!]
\includegraphics[scale=0.24]{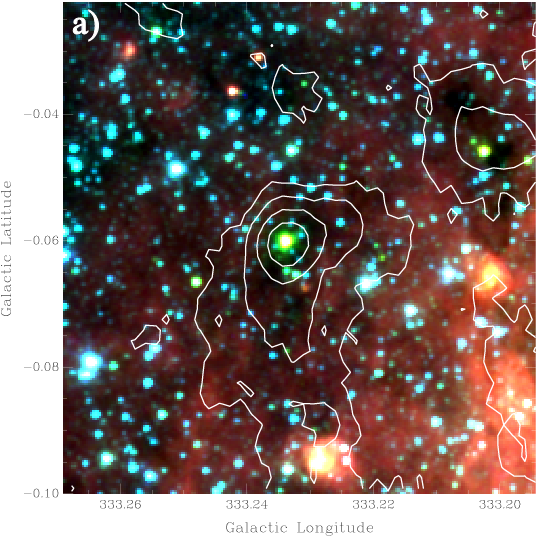}\includegraphics[scale=0.24]{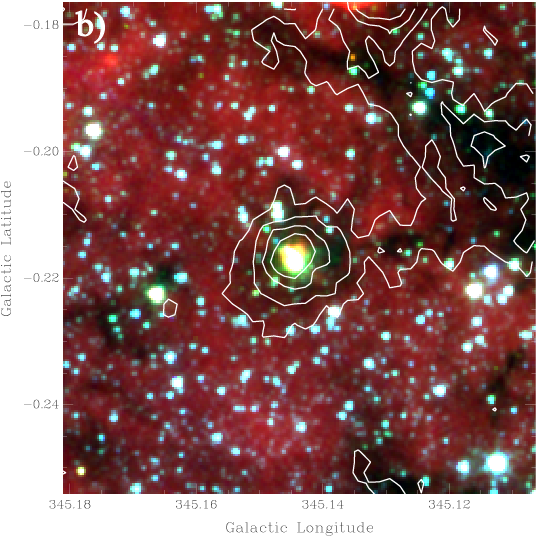}\includegraphics[scale=0.24]{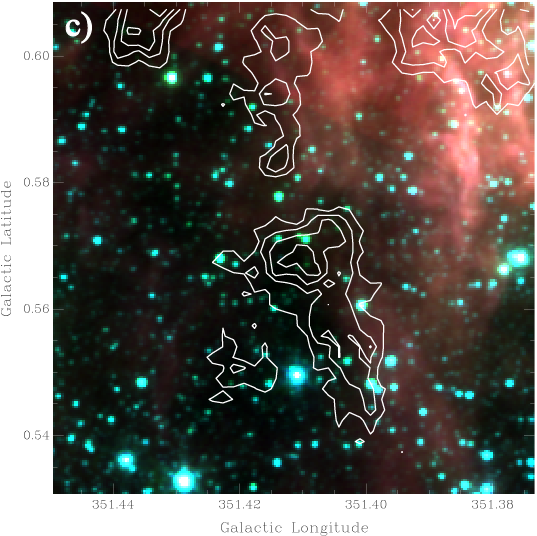}\includegraphics[scale=0.24]{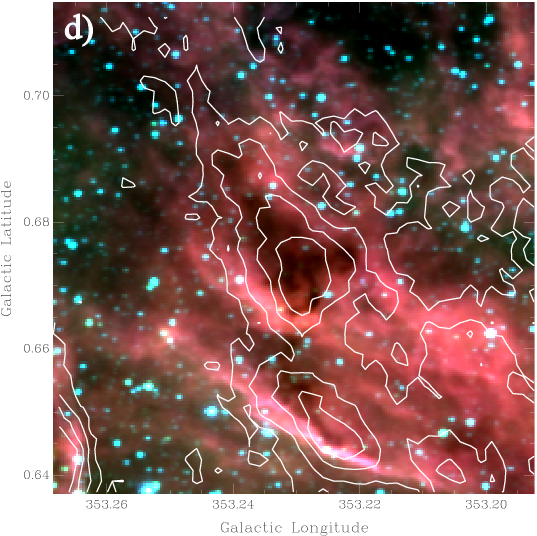}
\caption{$Spitzer$ three-color image (3.6~$\mu$m (blue), 4.8~$\mu$m (green), and 8.0~$\mu$m (red)) of the four ATCA sources observed in this paper overlaid with contours from APEX ATLASGAL 850~$\mu$m continuum observations. a) G333.234--00.061, with contours shown for [3, 10, 30, 70]~$\times$~$\sigma$, where $\sigma$ = 54~mJy$\,$beam$^{-1}$; b) G345.144--00.216, with contours shown for [3, 6, 10, 14]~$\times$~$\sigma$, where $\sigma$ = 58~mJy$\,$beam$^{-1}$; c) G351.409+00.567, with contours shown for [3, 5, 7, 9]~$\times$~$\sigma$, where $\sigma$ = 72~mJy$\,$beam$^{-1}$; and d) G353.229+00.672, with contours shown for [3, 6, 10, 14]~$\times$~$\sigma$, where $\sigma$ = 61~mJy$\,$beam$^{-1}$. 
\label{ATLASGALfigs} 
}
\end{figure*}

These sources represent some of the most extreme values of the [I(N$_2$H$^+$)/I(HCO$^+$)] ratio and have some of the brightest spectral lines of the anomalous sources. The two N$_2$H$^+$ rich sources are shown in Figure \ref{richanomalies}, and the two N$_2$H$^+$ poor sources are shown in Figure \ref{pooranomalies}. The two N$_2$H$^+$ rich sources each contain a bright region of enhanced 4.5 $\mu$m emission (a ``green fuzzy", Figure \ref{ATLASGALfigs}a, \ref{ATLASGALfigs}b) and compact 24~$\mu$m emission (Figure \ref{richanomalies}) indicative of a protostellar source.  The two N$_2$H$^+$ poor sources are dark in all bands (Figure \ref{pooranomalies}, \ref{ATLASGALfigs}c, \ref{ATLASGALfigs}d). Thus, contrary to prediction, the N$_2$H$^+$ rich sources contain compact heated regions, which should produce HCO$^+$ and destroy N$_2$H$^+$, while the N$_2$H$^+$ poor sources are dark, which could indicate a cold source that should be rich in N$_2$H$^+$. Spectral energy distribution fits to the $Herschel$ Hi-GAL data suggest that these clumps have masses larger than 200~$M_\sun$ (Y. Contreras, in preparation) and therefore are likely to produce at least one high-mass star $>$8~$M_\sun$ \citep{Jackson2013}.

Kinematic distances based on the MALT90 molecular velocities for G333.234--00.061 and G345.144--00.216 are $4.76^{+0.29}_{-0.29}$ and $1.86^{+0.68}_{-0.57}$~kpc respectively (S. Whitaker et al., in preparation). The other two dark clumps, G351.409+00.567 and G353.229+00.672, are poorly defined by kinematic distance because they lie at longitudes close to the Galactic center. However, G351.409+00.567 and G353.229+00.672 are coincident, both spatially and in velocity, with the star-forming complexes NGC~6334 and NGC~6357, respectively, which are reported in \citet{Russeil2012} to have distances of $1.7\pm0.3$~kpc and $1.9\pm0.4$~kpc. In this paper, we adopt these distances.

Following \citet{Hoq2013}, the dust temperature and mass column density for each clump can be estimated from the $Herschel$/Hi-GAL Survey (A. Guzm\'{a}n et al. in preparation). These parameters were derived by fitting a gray body model with a single temperature to the $Herschel$ Hi-GAL 160, 250, 350, and 500~$\mu$m maps. Each map was smoothed to the 500~$\mu$m resolution, allowing for the dust temperature estimates to be applicable on the scale of 36$\arcsec$. The central pixel values of the temperature maps for G333.234--00.061, G345.144--00.216,  G351.409+00.567, and G353.229+00.672 are 21, 18, 22, and 30~K respectively, and the central pixel for the mass column density is 0.75, 0.17, 0.074, and 0.075~g~cm$^{-2}$, respectively (A. Guzm\'{a}n, private communication). Using the distances for each source as discussed above, this suggests masses within the 36$\arcsec$ $Herschel$ beam of approximately 3100, 110, 35, and 45~$M_\sun$ for G333.234--00.061, G345.144--00.216,  G351.409+00.567, and G353.229+00.672 respectively. The clump for each source is extended and much larger than this size (as seen by the ATLASGAL emission in Figure \ref{ATLASGALfigs}), and thus is much more massive; these mass estimates simply give the approximate masses within the ATCA 3~mm primary beam. Moreover, $Herschel$ wavelengths are primarily sensitive to the cold component of gas, and thus a hot compact core (as probed by ATCA) that is much less massive than the cold clump mass can be completely hidden in the $Herschel$ bands.

For the two \nthp\ rich sources, G333.234--00.061 and G345.144--00.216, $Spitzer$ IRAC identifies protostars that are likely the hottest sources within the dark clump, suggesting the temperatures of 21 and 18~K, respectively, are lower limits to the core temperatures. The other two sources, G351.409+00.567 and G353.229+00.672, have no obvious protostellar sources, and thus the cores in these clumps may be much colder since dense environments can shield from external radiation.



\subsection{Observations and Data Reduction}
Using the ATCA we observed four molecular lines, HCO$^+$(1--0), HNC(1--0), $^{13}$CS(2--1), and N$_2$H$^+$(1--0), and two 2-GHz wide continuum bands (at 90 and 93~GHz) over the course of four days: from 24 August to 27 August 2013. The array configuration was H168, which provided an angular resolution of $\sim$2.2$\arcsec$ at these frequencies. The Compact Array Broadband Backend correlator setup for the observations was CFB 64M-32k, which provides 64~MHz zoom bands (i.e., a velocity range of $\sim$200~km$\,$s$^{-1}$) with 32~khz (0.11~km$\,$s$^{-1}$) spectral resolution along with two bands of 2~GHz bandwidth of continuum. For data reduction, we used MIRIAD \citep{Sault1995} and reduced the data in the standard manner. All four sources were observed every day, and the data from each day were combined. The phase calibrator used for G333.234--00.061 and G345.144--00.216 was 1600--44 and for G351.409+00.567 and G353.229+00.672 was 1742--289. For bandpass calibration, 1921--293 was used for 3 of the 4 days, while 1253--055 was used for the other day. Uranus was used as a flux calibrator, and the flux calibration is expected to be accurate within $\sim$20\%. Rest frequencies for HCO$^+$(1--0), HNC(1--0), and $^{13}$CS(2--1) were adopted from the Splatalogue database\footnote{Spectral line data were taken from the Spectral Line Atlas of Interstellar Molecules (SLAIM) (Available at http://www.splatalogue.net). (F. J. Lovas, private communication, \citealt{Remijan2007}).} and were taken to be 89.188526, 90.663572, and 92.49427~GHz respectively. A more accurate N$_2$H$^+$(1--0) rest frequency of 93.173772~GHz was adopted from \citet{Daniel2006}. The continuum observations were combined, resulting in a center frequency of 91.5~GHz.



The shortest baseline for this configuration is 61~m, which, at 91.5~GHz, makes these observations insensitive to extended emission larger than about 11$\arcsec$. Since these observations are only sensitive to structure smaller than 11$\arcsec$ and the sources have extended emission for the brightest molecular lines (i.e., HCO$^+$, HNC, and N$_2$H$^+$), very large negative sidelobes are evident in our images. Trying to suppress such large sidelobes proved to be futile. Instead, we generated ``dirty" images using natural weighting and identified the brightest emission peaks.  These peak intensity regions were then ``cleaned" using MIRIAD's \texttt{clean} algorithm.

Continuum emission was detected toward the two protostellar sources, G333.234--00.061 and G345.144--00.216; this emission was subtracted from the molecular line observations. All ATCA fluxes reported in this paper have been corrected for primary beam attenuation.

Also discussed in this paper are 22.2 GHz continuum observations of G333.234--00.061 taken with the ATCA on 5 May 2014 in the 1.5D array configuration. These data were reduced in a similar manner as the 91.5 GHz observations, using 1253--055 as the bandpass calibrator, 1646--50 as the phase calibrator, and 1934--638 as the flux calibrator. Fluxes at 22.2 GHz are accurate to within $\sim$10\%.

Additionally for these four sources, we present the spectra from the MALT90 survey (J. Rathborne et al. in preparation). These lines are summarized in Table \ref{tab:m90}.

\begin{deluxetable}{lc@{}c}
\tablecolumns{3}
\tabletypesize{\small}
\tablewidth{0pt}
\tablecaption{MALT90 Spectral Lines \label{tab:m90}}
\tablehead{\colhead{Species} & \colhead{Main transition} & \colhead{Frequency (GHz)}
}
\startdata
{\bf \nthp} & {\bf \emph{J} = 1 -- 0} & {\bf 93.173772} \\
 {\bf $^{13}$CS} &  {\bf \emph{J} = 2 -- 1} &  {\bf 92.494303} \\
H & 41$\alpha$ & 92.034475 \\
CH$_3$CN & $J_K$ = 5$_1$ -- 4$_1$ &  91.985313\\ 
HC$_3$N & $J$ = 10 -- 9 & 90.979020 \\  
$^{13}$C$^{34}$S & $J$ = 2 -- 1 & 90.926036 \\
{\bf HNC} & {\bf \emph{J} = 1 -- 0} & {\bf 90.663572} \\
HC$^{13}$CCN & $J$ = 10 -- 9, $F$ = 9 -- 8 & 90.593059 \\
{\bf HCO$^+$} & {\bf \emph{J} = 1 -- 0} & {\bf 89.188526} \\
HCN & $J$ = 1 -- 0 & 88.631847 \\
HNCO & $J_{K_a,K_b}$ = 4$_{0,4}$ -- 3$_{0,3}$ & 88.239027 \\
HNCO & $J_{K_a,K_b}$ = 4$_{1,3}$ -- 3$_{1,2}$ & 87.925238 \\
C$_2$H & $N$ = 1 -- 0, $J$ = $\frac{3}{2} - \frac{1}{2}$, & 87.316925 \\
& $F$ = 2 -- 1 & \\
HN$^{13}$C & $J$ = 1 -- 0 & 87.090859 \\
SiO & $J$ = 2 -- 1 & 86.847010 \\
H$^{13}$CO$^+$ & $J$ = 1 -- 0 & 86.754330

\enddata
\tablecomments{Bold rows indicate spectral lines that were also observed with ATCA for the four sources in this paper.}
\end{deluxetable}

\section{N$_2$H$^+$ Rich Sources}\label{sec:n2hprich}

\subsection{G333.234--00.061}\label{sec:G333}

\subsubsection{Continuum}\label{sec:g333_cont}
\begin{figure}[ht!] Ê
\begin{center}
\includegraphics[scale=0.3]{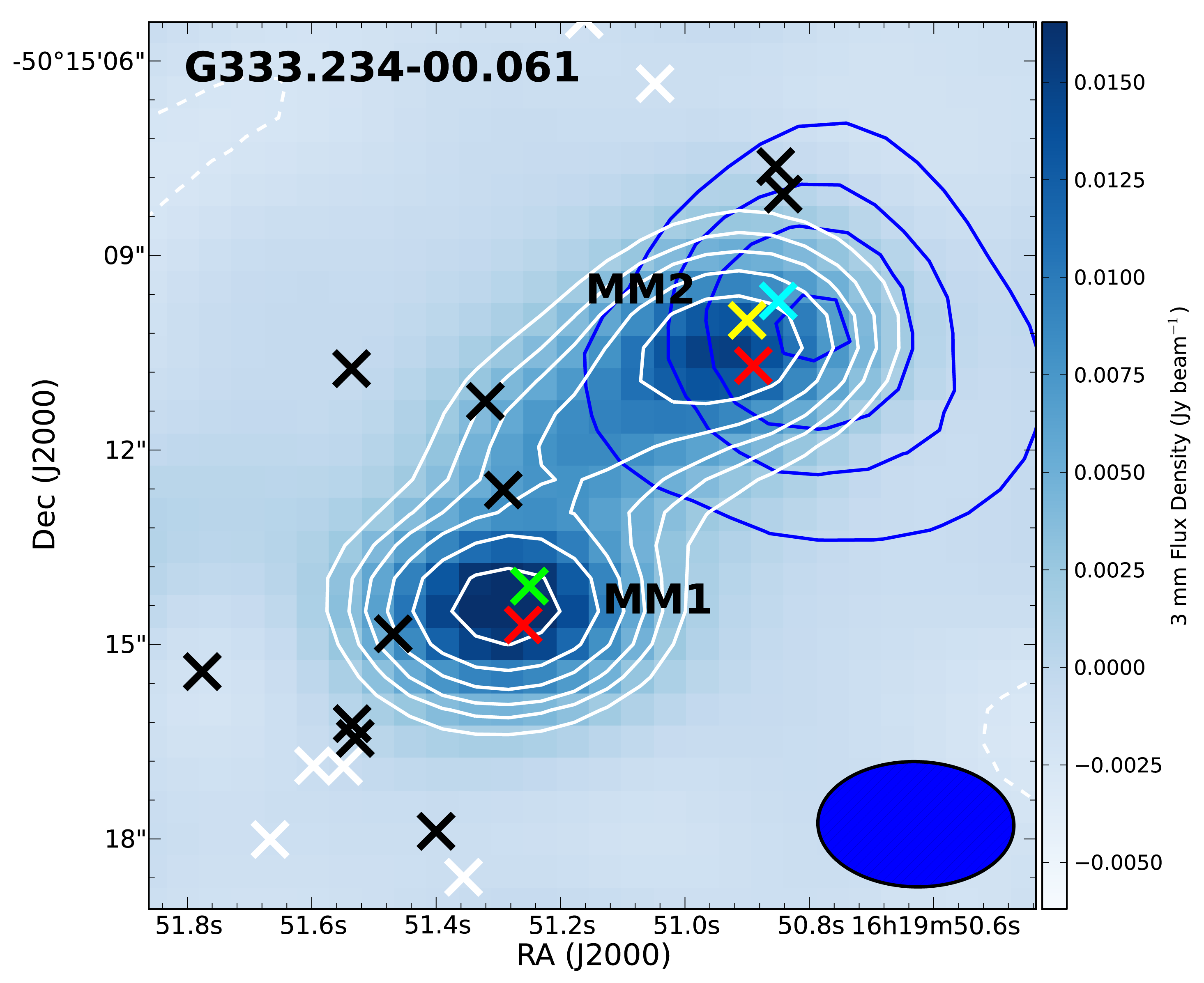}
\caption{The ATCA 3.3~mm continuum image for G333.234--00.061, corrected for the primary beam. White contours and color-scale are shown for the 3.3~mm continuum, with contours shown for [--3, 3, 5, 7, 10, 14, 20]~$\times$~$\sigma_{3.3\rm{mm}}$, where $\sigma_{3.3\rm{mm}}=0.17$~mJy$\,$beam$^{-1}$. Blue contours are the $Spitzer$ IRAC 8.0~$\mu$m with contour levels [200, 300, 500, 1000]~$\times$~1 MJy$\,$sr$^{-1}$. The white, black, and green $\times$ marks show locations of class I methanol masers for 36, 44 \citep{Voronkov2014}, and 95.1~GHz \citep{Ellingsen2005}, respectively. The red, cyan, and yellow $\times$ marks indicate 6.7~GHz class II methanol \citep{Caswell2011}, 22 GHz water \citep{Breen2010}, and 1.7~GHz OH masers \citep{Caswell1998}, respectively. \label{g333_cont}
}
\end{center}
\end{figure}

The continuum map for G333.234--00.061 is shown in Figure \ref{g333_cont} and shows two very bright 3.3~mm continuum cores in the field. The southern core (henceforth MM1) is significantly brighter than the northern core (henceforth MM2). A weaker third continuum core may be present between the two main continuum peaks. Neither MM1 or MM2 is resolved. The continuum observations have a synthesized beam size of 2$\farcs$4 (calculated as the geometric mean of the synthesized beam's full width at half maximum (FWHM) major and minor axes), corresponding to an upper limit for the core sizes of $\sim$0.06~pc.

We used MIRIAD task \texttt{maxfit} to locate the intensity maxima; the results place MM1 at R.A. (J2000.0$)=16^{\rm{h}}19^{\rm{m}}51\fs29$ and decl.~$=-50\arcdeg15\arcmin14\farcs4$ and MM2 at R.A. (J2000.0$)=16^{\rm{h}}19^{\rm{m}}50\fs94$ and decl.~$=-50\arcdeg15\arcmin10\farcs5$. These peaks are separated by 5$\farcs$1 or a projected distance of $0.12\pm0.012$~pc (where the error only comes from the uncertainty in distance). Since these cores have similar velocities (as discussed in Section \ref{sec:g333_lines}) and are associated with the same MALT90 clump, their separation is probably not much larger than this projected distance. The clump velocity is a better indicator of the kinematic distances than the actual core velocities because MM1 and MM2 may have peculiar velocities within the clump. If we were to assume the projected distance is the same as the radial distance, the separation of the cores would increase to 0.17~pc. However, since we are unable to determine the radial distance with the given data, we will simply refer to the separation as a ``projected distance of 0.12~pc" in the rest of the paper.

MM2 is associated with $Spitzer$ IRAC emission in all four bands (only 8.0~$\mu$m is shown in Figure \ref{g333_cont}) and with $Spitzer$ MIPS at 24~$\mu$m. MM1 is not detected in any of the IRAC bands. The resolution for MIPS at 24~$\mu$m ($\sim$6$\arcsec$) is insufficient to resolve the two cores (separated by 5$\farcs$1), though the 24~$\mu$m emission is centered on MM2 and is not elongated toward MM1. This 24~$\mu$m compact emission and presence of a green fuzzy (Figures \ref{richanomalies} and Figure \ref{ATLASGALfigs}) indicates that MM2 is likely protostellar and MM1 is colder and at a younger age; however, MM1 is likely protostellar as well since it is coincident with a class II (radiatively pumped) methanol maser, as discussed in Section \ref{sec:masers}.


Since MM1 and MM2 are unresolved, we approximate the integrated flux densities to be the same as the peak flux densities integrated about the synthesized beam, giving integrated flux densities of 19.0 and 15.1~mJy for MM1 and MM2 respectively. The total flux density of the continuum is 47.5 mJy (calculated by summing the flux in a polygon around the emission), giving a leftover integrated flux density (which is dominated by the connecting dust lane) of 13.4~mJy.




To calculate the dust masses of the cores, we use the standard equation from \citet{Hildebrand1983}
\begin{equation}
M_D=\frac{F_\nu^iD^2}{\kappa_\nu B_\nu(T_D)}, \label{h83}
\end{equation}
where $F_\nu^i$ is the observed integrated flux density, $D$ is the distance to the source, $\kappa_\nu$ is the dust opacity, and $B_\nu(T_D)$ is the Planck function given at dust temperature $T_D$. As discussed in Section \ref{sec:classsourceselection}, $Herschel$ data show that the cold clump mass in the ATCA beam is large, so the derived dust temperature at the clump-scale (36$\arcsec$ resolution) is 21~K and is likely a lower limit of $T_D$ for these protostellar sources. The 3~mm continuum emission is not typically dominated by the clump or the central protostar but rather by the thermal dust emission from the surrounding dust cocoon at a radius of $\sim$1000 to 10000 AU. Dust temperatures in a typical high-mass hot molecular core at these radii vary from $\sim$30~K to 120~K \citep{Osorio1999}. For this paper, we adopt $T_D=100$~K. This temperature is likely high for MM1 since the lack of IRAC emission indicates it may be much colder than MM2. Assuming that $\kappa_\nu \propto \nu^\beta$ and $\beta$, the dust emissivity index, is 1.5, we scaled $\kappa_{1.3\,\rm{mm}} = 0.90$~cm$^2$~g$^{-1}$ \citep[][assuming thin ice mantles and gas density of 10$^6$~cm$^{-3}$]{Ossenkopf1994} to $\kappa_{3.3\,\rm{mm}} = 0.22$~cm$^2$~g$^{-1}$. We assume a gas-to-dust mass ratio (GDR) of 100 to estimate the core mass. From these values, we calculate gas masses of 36~$M_\sun$ and 29~$M_\sun$ for MM1 and MM2 respectively. The remaining material of the complex at this temperature has a mass of 26~$M_\sun$, suggesting the total mass of the core complex to be $\sim$100~$M_\sun$.

Estimating errors of these masses is difficult since the parameters are highly uncertain. If our assumed values are the most probable, we can estimate errors for our sources. We assume a 20\% error in $F_{3.3}^i$ and  $\kappa_{1.3\,\rm{mm}}$. As discussed in Section \ref{sec:classsourceselection}, the kinematic distance with uncertainties is $D=4.76^{+0.29}_{-0.29}$. For the errors of the other parameters, we assume they are uniformly distributed in the following ranges: GDR~$\in (70, 150)$, $\beta \in (1,2)$, and $T \in ($20~K, 120~K). Our mass estimates with errors for MM1 and MM2 are thus 36$^{+38}_{-15}$~$M_\sun$ and 29$^{+30}_{-12}$~$M_\sun$. However, our parameters were selected because they give conservatively low mass estimates. For example, if we change the $T_D$, one of the most dependent variables for estimating the mass, from 100~K to 50~K (with the same range for each parameter), the masses would be 74$^{+58}_{-46}$ and 59$^{+47}_{-36}$~$M_\sun$ respectively. For the rest of this section we will use the warmer masses of 36~$M_\sun$ and 29~$M_\sun$ for MM1 and MM2. Given that the parameters for Equation \ref{h83} were chosen to give conservatively low masses, these masses are likely lower limits. The ratio between the masses of MM1 and MM2, however, is much less uncertain since all parameters in Equation \ref{h83} are likely similar for both sources except $F_\nu^i$ and possibly $T_D$. Therefore, MM1 is at least 25\% more massive than MM2 since MM1's flux is 25\% higher and it is likely at a colder temperature.

%


In the infrared dark cloud SDC335.579-0.272, \citet{Peretto2013} found the highest 3~mm luminosity ever seen for a protostellar core ($F_{3.2 \rm{mm}}^i$~=~101 mJy at a distance of 3.25~kpc) for a core diameter $<$0.1~pc. Assuming $T_D=50$~K and $\kappa_{3.2\,\rm{mm}} = 0.087$~cm$^2$~g$^{-1}$ (using $\beta=2$), \citeauthor{Peretto2013} estimated a core mass of 545~$M_\sun$. Indeed, if we assume the same dust opacity and temperatures for MM1 and MM2, both of these cores would have masses around 200~$M_\sun$. \citeauthor{Peretto2013} compared masses for about 30 protostellar cores, standardizing their masses with the same dust opacity law; such a comparison suggests that MM1 and MM2 are perhaps the second and third most massive protostellar cores known for diameters $<$0.1~pc. Regardless of the exact values of $\kappa$ and $T_D$, MM1 and MM2 are two of the most massive protostellar cores known, and they are forming in close proximity to each other. In this paper, we have assumed a higher $T_D$ and lower $\beta$ compared to most of the studies mentioned in \citeauthor{Peretto2013}, which lowers our mass estimates.
 
 

For these cores we calculated the beam-averaged H$_2$ column density, $N$(H$_2$), using
\begin{equation}
N(\mbox{H}_2)= \frac{F_\nu^p}{B_\nu(T_D) \mu m_p \kappa_{\nu}\Omega_b}, \label{cd}
\end{equation}
where $F_\nu^p$ is the peak flux density, $m_p$ is the mass of a proton, $\mu$ is the mean molecular weight per hydrogen molecule, and $\Omega_b$ is the solid angle subtended by the beam. Given $F_\nu^p$ of 19.0 and 15.1~mJy$\,$beam$^{-1}$ for MM1 and MM2, respectively, and assuming $\mu=2.8$ \citep{Kauffmann2008} and a gas-to-dust mass ratio of 100, we determine a $N$(H$_2$) of $4.6\times10^{23}$ and $3.7\times10^{23}$~cm$^{-2}$ for MM1 and MM2.

We calculated the mean volume density of MM1 and MM2 using the expression
\begin{equation}
n(\mbox{H}_2)=\frac{M_G}{\mu m_p V},
\end{equation}
given the gas mass $M_G$, mass of a proton $m_p$, and an assumed spherical volume $V=\frac{4}{3} \pi r^3$, where $r$ is the radius of the core. Since the cores are unresolved, we use a radius of half the synthesized beam (i.e., a radius of 0.03~pc), which provides a lower limit of the mean volume density. The $n$(H$_2$) for MM1 and MM2 is $4.6\times10^6$ and $3.7\times10^6$~cm$^{-3}$ respectively.

Thus far, we assumed that all of the 3.3 mm continuum flux arises from thermal dust emission and have neglected any contamination from free-free emission. Additional ATCA observations at 22.2~GHz (1.35~cm) detected both MM1 and MM2. These cores are resolved with a synthesized beam of $1\farcs33\times1\farcs07$ ($\sim$0.03 pc resolution) and have peak fluxes of 0.30 and 0.39 mJy$\,$beam$^{-1}$ for MM1 and MM2 respectively. If we smooth these observations to the same resolution as the 91.5 GHz continuum observations, the peak fluxes are 0.66 mJy$\,$beam$^{-1}$ and 0.70 mJy$\,$beam$^{-1}$ for MM1 and MM2 respectively. Assuming that flux scales as $F_\nu \propto \nu^{\alpha}$, the spectral index $\alpha$ is 2.4 and 2.2 for MM1 and MM2 respectively. Although optically thick free-free emission can have spectral indices up to two, typically hypercompact (HC) \ion{H}{2} regions \citep[sizes less than $\sim$0.03~pc,][]{Franco2000} have $\alpha \approx 1$ \citep[e.g.,][]{Cyganowski2011} which may be due to non-uniform gas density \citep{Hoare2007}. Dust emission from high-mass protostellar objects has a spectral index $\alpha \geq 2$, with typical values of 3 to 4.5. Toward MM1 and MM2, thermal dust emission is certainly significant since $\alpha$ is larger than 2, but $\alpha$ is lower than expected for gray body dust emission at a single dust temperature.

If we make the conservative assumption that all the 91.5 GHz continuum emission is from free-free emission, the brightness temperature of the free-free emission can be calculated via
\begin{equation}
T_b = \frac{F_\nu c^2}{2\nu^2k\Omega_b} = \phi T_e(1-e^{-\tau_\nu}),
\end{equation}
where $\Omega_b$ is the solid angle subtended by the synthesized beam (3\farcs02 $\times$ 1\farcs93), $\phi$ is the beam filling factor, $T_e$ is the electron temperature, and $\tau_\nu$ is the optical depth. Although MM1 and MM2 could be optically thin at 91.5~GHz, $\alpha$ is 2.4 and 2.2 for MM1 and MM2 respectively. If free-free emission indeed dominates the 3.3 mm continuum flux, these spectral indices can only be explained by optically thick free-free emission. For optically thick free-free emission, the beam filling factor is therefore $\phi = T_b/T_e = \Omega_s/\Omega_b$, where $\Omega_s$ is the solid angle subtended by the source. Although $T_e=10^4$~K is typically assumed in literature, electron temperatures in the Galaxy have been reported to be as low as $\sim$5000~K \citep[e.g.,][]{Quireza2006}, and we have used this conservative value for our calculation to maximize $\phi$. The measured $T_b$ from the peak flux densities at 91.5~GHz are 0.48 and 0.38~K for MM1 and MM2 respectively. For optically-thick free-free emission, this results in a maximum $\phi$ of 9.5$\times10^{-5}$ and 7.6$\times10^{-5}$ for MM1 and MM2, suggesting that the sources have diameters of 0$\farcs$024 and 0$\farcs$021 respectively. This indicates a very small $maximum$ size of $\sim$100 AU for optically thick free-free emission. An \ion{H}{2} region this small likely only occurs for a brief period of a massive protostar's life -- this is extremely unlikely to be seen for two sources. Thus, we conclude that free-free dominant emission is unlikely and that most of our emission at 91.5~GHz comes from dust. The same analysis of the 22.2~GHz observations also finds a maximum size of $\sim$100~AU for an optically thick free-free source that dominates the observed emission. If we assume all the emission at 22.2~GHz is optically thin free-free emission with a flat spectral index ($\alpha=0$), it would contribute less than 5\% to the total flux at 91.5~GHz.


A mix of thermal dust emission and free-free emission is still possible. Observations in other star-forming regions have shown that although 91.5~GHz is dominated by dust, frequencies lower than $\sim$40~GHz can be dominated by a thermal jet \citep{Brooks2007} or free-free emission from a (HC) \ion{H}{2} region \citep{Zhang2014}, though a HC \ion{H}{2} region is not likely, as explained in the previous paragraph. \citet{Brooks2007} showed that when using only 25 and 88~GHz continuum observations, the spectral index of the young stellar object IRAS~16547--4247 is $\alpha = 1.6$. However, between 88~GHz and 250~GHz, $\alpha = 4.7$. Thus, the fact that MM1 and MM2 have spectral indices less than expected could simply be due to a mix between free-free emission from a thermal jet and dust emission. Since we only have two continuum points at high angular resolution, we cannot estimate the free-free contamination of the dust emission. However, since the 3.3~mm emission appears to be dominated by dust, we neglect other possible components.


\subsubsection{Molecular lines}\label{sec:g333_lines}
\begin{figure*} [ht!]Ê
\begin{center}
\includegraphics[scale=0.7]{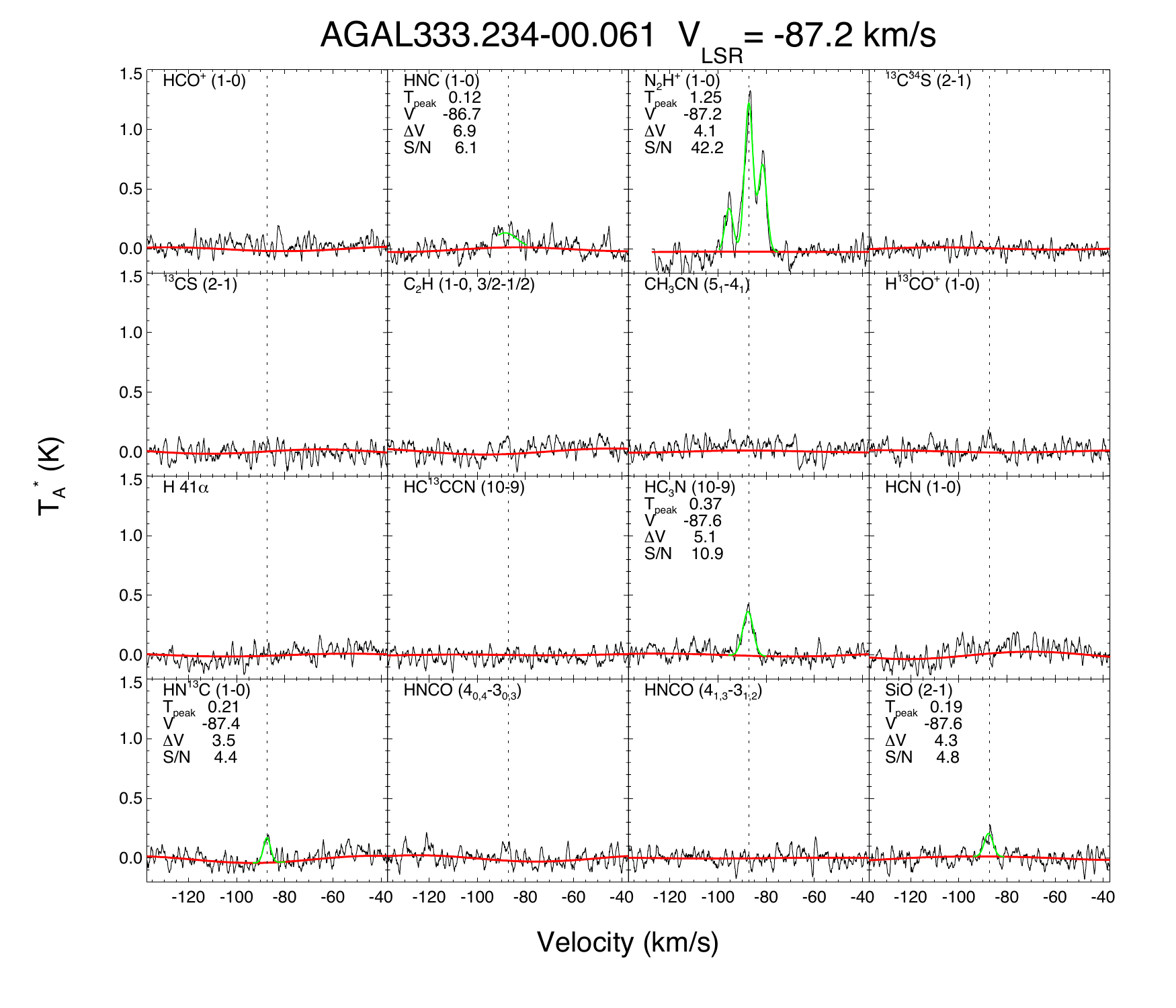}
\caption{MALT90 (resolution of 38$\arcsec$) observations of G333.234--00.061 for 16 observed lines (J. Rathborne et al., in preparation). The spectra have been smoothed using a Savitzky-Golay kernel. The consensus velocity, V$_{\mbox{\scriptsize LSR}}$, is indicated at the top of the figure and is shown as a reference line in each panel. Each panel shows different observed line transitions with the red curve showing the fit to the baseline and the green curve showing the fit to the line. The line fit is only shown if the signal-to-noise is larger than three, and the corresponding fit parameters are shown in the top-left of the corresponding panel. $T_{\mbox{\scriptsize peak}}$ is the peak line intensity in K, $V$ is the center velocity and $\Delta V$ is the linewidth in \kms, and S/N is the signal-to-noise of the line fit. \label{g333_malt90}
}
\end{center}
\end{figure*}

Figure \ref{g333_malt90} shows the 16 lines observed in the MALT90 survey (resolution of 38$\arcsec$) toward G333.234--00.061, which provide a comparison of the clump ($\sim$1~pc size scale) molecular properties to that of the cores ($<$0.1~pc size scale) probed by the ATCA. With the Mopra 22~m single dish, HCO$^+$(1--0) and $^{13}$CS(2--1)  are undetected, HNC(1--0) is marginally detected, and N$_2$H$^+$(1--0) is strongly detected. The N$_2$H$^+$ line is broad, indicating a large velocity dispersion or multiple velocities components. Many other lines are detected or marginally detected: C$_2$H(1--0), H$^{13}$CO$^+$(1--0), HC$_3$N(1--0), HN$^{13}$C(1--0), HNCO(4$_{0,4}$--3$_{0,4}$), and SiO(2--1). The fact that HCO$^+$ is undetected and HNC is marginally detected but H$^{13}$CO$^+$ and HN$^{13}$C are detected indicates that the optically thicker HCO$^+$ and HNC lines are likely self-absorbed.

\begin{figure*} [ht!]Ê
\begin{center}
\includegraphics[scale=0.075]{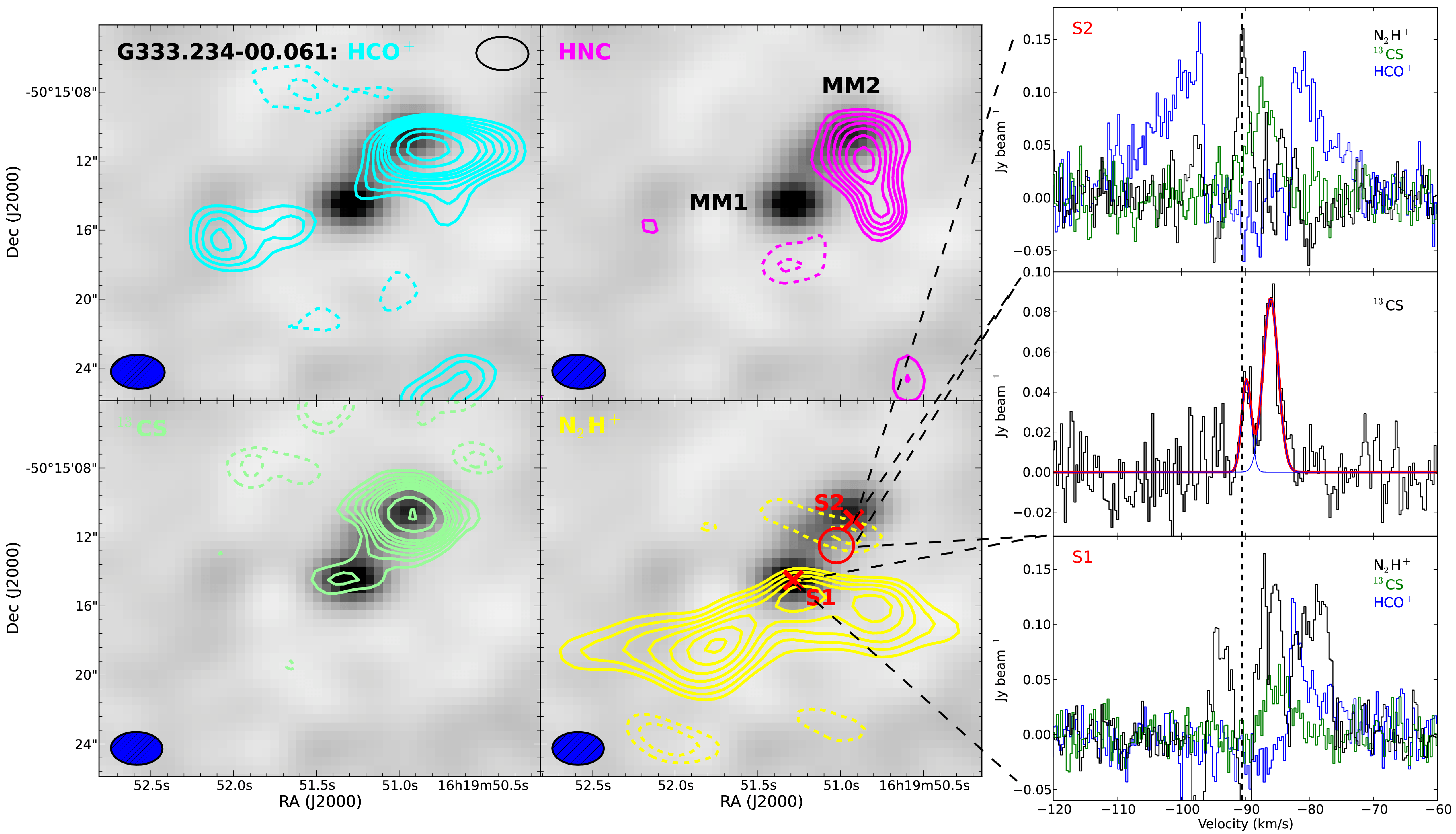}
\caption{G333.234--00.061 ATCA integrated intensity maps of the molecular lines HCO$^+$(1--0), HNC(1--0), $^{13}$CS(2--1), and N$_2$H$^+$(1--0) overlaid on a grayscale map of the 3.3~mm continuum.  Reference contours are shown for [--4, --3, 3, 4, 5, 6, 7, 8, 9, 12, 15]~$\times~f$ for $f_{\rm{HCO}^+} =  0.075$, $f_{\rm{HNC}} =  0.05$, $f_{^{13}\rm{CS}} =  0.022$, and $f_{\rm{N}_2\rm{H}^+} =  0.22$~Jy$\,$beam$^{-1}$~km$\,$s$^{-1}$. The open black beam (top panel, top right) represents the continuum resolution while the filled blue beam (bottom left) represents the panel's molecular line resolution. Lines were integrated based on the following velocity ranges: HCO$^+$: --113.2 to --72.0~km$\,$s$^{-1}$, HNC: --98.5 to --73.3~km$\,$s$^{-1}$, $^{13}$CS: --95.3 to --79.8~km$\,$s$^{-1}$, and \nnhp: --99.3 to --76.2~km$\,$s$^{-1}$. Selected spectra are shown on the right for the positions shown in the N$_2$H$^+$ map, with the red $\times$ marks showing spectra at the corresponding pixel and the red circle showing the area used to present an averaged $^{13}$CS spectrum. The spectra have been Gaussian smoothed and show a reference dashed line at a velocity of --90.5~\kms. The central $^{13}$CS spectrum is fit with two Gaussians (indicated in blue, with the sum in red). The redshifted (right) Gaussian has an amplitude $A = 87\pm7$~mJy\,beam$^{-1}$ (or $2.2\pm0.18$~K), central velocity $v=-86.0\pm0.12$~km$\,$s$^{-1}$, and linewidth $\Delta v=2.8\pm0.29$~km$\,$s$^{-1}$ and the blueshifted (left) Gaussian has $A = 46\pm9$~mJy\,beam$^{-1}$ (or $1.1\pm0.23$~K), $v=-89.8\pm0.18$~km$\,$s$^{-1}$, and $\Delta v=1.7\pm0.43$~km$\,$s$^{-1}$. 
\label{g333_lines}
}
\end{center}
\end{figure*}

Integrated intensity maps of the molecular lines observed with the ATCA are shown in Figure \ref{g333_lines}, with strong detections for all four of the observed molecular lines (HCO$^+$, HNC, $^{13}$CS, and N$_2$H$^+$). For the 3 strongest lines, HCO$^+$, HNC, and N$_2$H$^+$, the presence of extended emission has caused large sidelobes. Contours have been set in Figure \ref{g333_lines} to show minimal sidelobes, which causes the integrated intensity maps not to show some real emission (e.g., MM1 has definite detections for all lines). Thus, Figure \ref{g333_lines} only shows the locations of the strongest emission.

HCO$^+$ and HNC are especially strong for MM2 while the N$_2$H$^+$ is primarily concentrated in an elongated structure south of both cores. $^{13}$CS, the most optically thin of the probed high-density tracers, traces well the two continuum cores and is also especially strong toward MM2. 

The fact that $^{13}$CS is much brighter for the lower mass core MM2 is intriguing. Estimating the optical depth for $^{13}$CS without observations of other CS isotopologues is difficult, especially since the compact emission is not resolved. Therefore, we take an analytical approach in order to discern why $^{13}$CS is brighter for MM2. The critical density of $^{13}$CS(2--1) depends on temperature and is a few times $10^5$~cm$^{-3}$, which is well below the lower limit of the densities of the cores (few times $10^6$~cm$^{-3}$, Section \ref{sec:g333_cont}). Therefore the lines are thermalized for both cores, causing the excitation temperature, $T_{\rm{ex}}$, to be approximately equal to the gas kinetic temperature. The $^{13}$CS brightness temperature can be characterized via $T_b = \phi T_{\rm{ex}}(1-e^{-\tau_\nu})$. In the optically thin limit, $T_b = \phi T_{\rm{ex}} \tau_\nu$, and for the optically thick limit $T_b = \phi T_{\rm{ex}}$. Since $^{13}$CS is brightest for MM2, the $^{13}$CS emission for MM1, when compared to MM2, (1) has a lower optical depth (in the optically thin limit), (2) has a lower excitation temperature, (3) has more self-absorption, and/or (4) comes from a smaller region (i.e., has a smaller $\phi$). Since column densities as calculated by dust (Section \ref{sec:g333_cont}) are higher than MM1, (1) would suggest that in the optically thin limit, $^{13}$CS in MM2 must be more abundant unless the MM2 excitation temperature is much higher. (2) and (3) would both suggest that MM1 is colder than MM2. (4) suggests that MM1 is more compact. Thus, this analysis suggests that MM1 could be colder and/or more compact.


Figure \ref{g333_lines} (right) shows selected spectra for the molecular lines HCO$^+$, $^{13}$CS, and N$_2$H$^+$ for MM1 (S1, at the continuum peak), MM2 (S2, at a location to best display the outflow wings), and the average spectrum of a small area near S2. The N$_2$H$^+$ spectra are contaminated by negative sidelobes (this is particularly evident in S1), causing the fluxes to be inaccurate. 

%

The N$_2$H$^+$ spectrum at S1 (Figure \ref{g333_lines}, bottom right) suggests two velocity components, though careful analysis of the spectral cube indicates that a negative sidelobe lies between these two ``components" which may have erroneously created the two peaks. Since $^{13}$CS is optically thinner than N$_2$H$^+$, $^{13}$CS is more likely to trace the core while N$_2$H$^+$ is more likely to trace the clump. The fact that $^{13}$CS is slightly redder than the N$_2$H$^+$ line may indicate that the core is moving relative to the clump. A single Gaussian fit to the $^{13}$CS line gives an amplitude $A = 45\pm5$~mJy\,beam$^{-1}$ (or 1.1$\pm$0.14~K), a central velocity $v=-84.7\pm0.27$~km$\,$s$^{-1}$, and velocity linewdith $\Delta v=4.6\pm0.63$~km$\,$s$^{-1}$.

The HCO$^+$ spectrum is extremely self-absorbed in both locations. Emission at the systemic velocity is absent and only redshifted emission is detected. The HCO$^+$ spectrum (as well as HNC, not shown) also shows absorption against the continuum at the locations of the self-absorption. Although it is difficult to measure the absorption against the continuum due to the sensitivity of the observations, the depth of the continuum absorption is comparable to the peak of MM1's continuum emission of 18.9~mJy\,beam$^{-1}$. Thus, the HCO$^+$ line is optically thick.

The spectra for S2 (Figure \ref{g333_lines}, top right) show two broad HCO$^+$ outflow wings with a deep self-absorption feature splitting the red and blue wings. N$_2$H$^+$ and $^{13}$CS emission peaks are between the HCO$^+$ outflow wings at similar velocities of the HCO$^+$ self-absorption feature. Similar to S1, the S2 velocity offset between $^{13}$CS and N$_2$H$^+$ may indicate that the core is moving relative to the clump. Moreover, there is also HCO$^+$ absorption against the continuum between the outflow wings, approximately at MM2's continuum peak (13.6~mJy\,beam$^{-1}$). The $^{13}$CS and N$_2$H$^+$ spectra show marginal evidence for two velocity components. A $^{13}$CS spectrum averaged over the circle just southeast of MM2 is shown in the middle right panel of Figure \ref{g333_lines}.  The two velocity components were fitted with two Gaussians. The velocity of the brighter component is $v=-86.0$~km$\,$s$^{-1}$, while that of the fainter component is $v=-89.8$~km$\,$s$^{-1}$. Since the optical depth should be larger at positions approaching MM2, and the two distinct Gaussian components become more obvious at positions displaced from MM2, two velocity components seem to be a more probable explanation of the double $^{13}$CS spectral peaks rather than self-absorption in this transition.


Although the lack of HCO$^+$ emission at the systemic velocity of MM1 could conceivably be explained not by self-absorption but rather by the interferometer resolving out large scale emission, we tested this idea and found that it is unlikely. If we smooth the ATCA observations to the size of the Mopra beam, the resulting HCO$^+$ line profile (which is not Gaussian due to the self-absorption feature) has a peak flux of 0.07~K and a velocity of --82.3~km$\,$s$^{-1}$. This brightness temperature is roughly the same as the rms noise for the smoothed Mopra spectrum in Figure \ref{g333_malt90}. If the lack of HCO$^+$ emission was indeed a case of resolving out structure, the peak brightness of the missing flux would be much higher than 0.07~K and would be spatially distributed at scales larger than 11$\arcsec$ (the largest structure the ATCA observations are sensitive to); thus, the single dish brightness would be much higher than 0.07~K and would be easily detected by Mopra. Moreover, the fact that there is HCO$^+$ absorption against the continuum and that N$_2$H$^{+}$ does not show the typical 1:5:3 intensity ratios between the three main hyperfine peaks shows that the central region is optically thick, conducive to self-absorption. Also, as already discussed, the single-dish detections of isotopologues (e.g., HN$^{13}$C, H$^{13}$CO$^+$) at intensity ratios much higher than the $^{13}$C/$^{12}$C abundance ratios demonstrate very large optical depths. Therefore, these observations in conjunction with the MALT90 observations indicate that the HCO$^+$ is deeply self-absorbed. 

%
%
The S2 self-absorption feature occurs over a velocity space of 14~\kms~(--97 to --83~\kms, Figure \ref{g333_lines}, top right), and the flux of the ATCA observations is approximately zero in the entire primary beam within this velocity range (not shown). This is an exceptionally broad for an absorption feature that is spatially distributed throughout the clump. In order to create such an absorption feature, there must be optically thick cold gas over a velocity range of 14~\kms. Possible mechanisms to produce such a broad absorption line include infall of cold gas on the clump scale or extreme saturation of a very thick line.

\begin{figure} [ht!]Ê
\begin{center}
\includegraphics[scale=0.28]{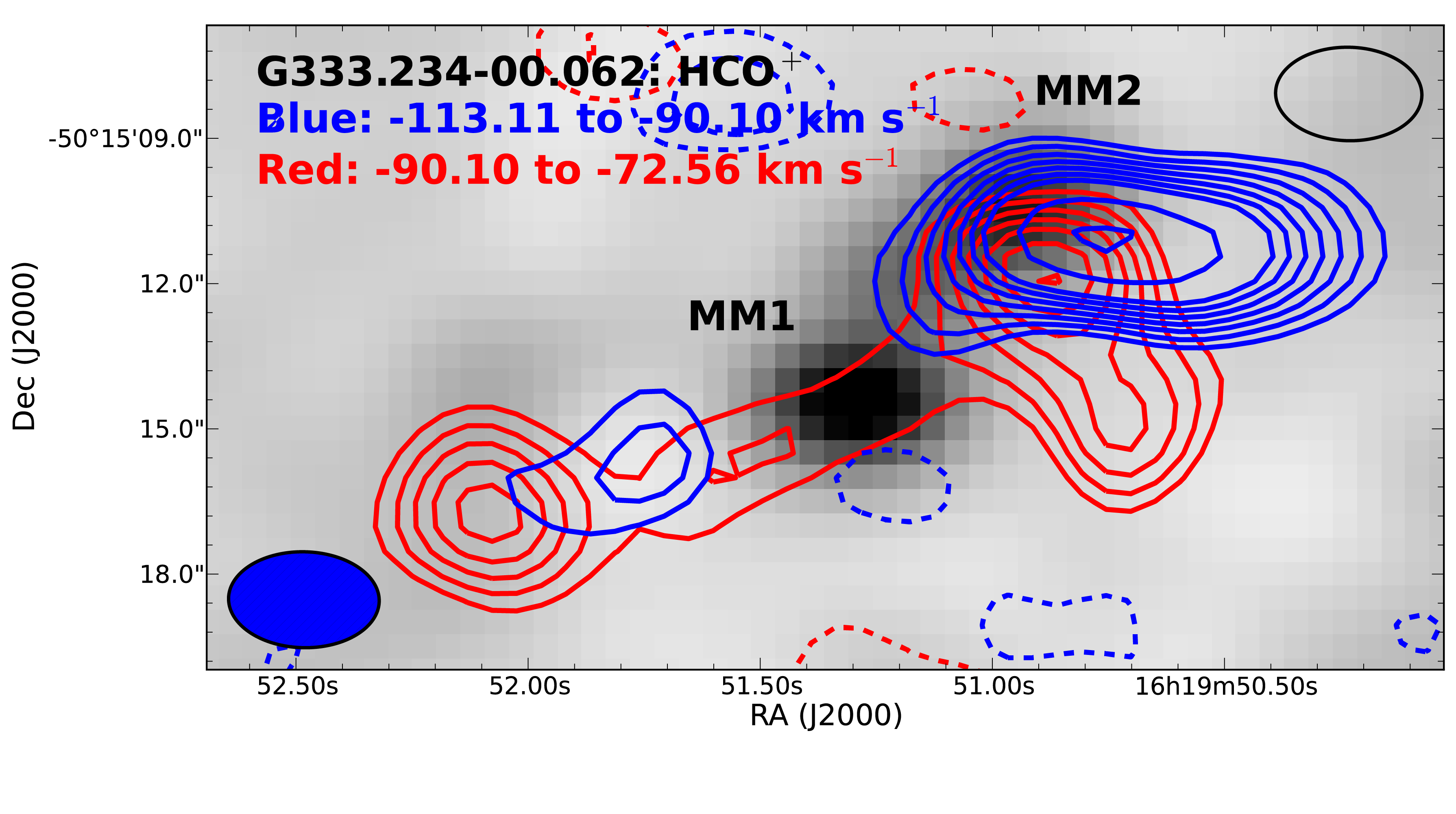}
\caption{The ATCA HCO$^+$(1--0) integrated intensity observations showing the red and blue components of G333.234--00.061. MM2 has a strong outflow almost along the line of sight. Contours are shown for [--4, --3, 3, 4, 5, 6, 7, 8, 9, 12, 15]~$\times$~0.065~Jy$\,$beam$^{-1}$~km$\,$s$^{-1}$. The open black beam (top right) represents the continuum resolution while the filled blue beam (bottom left) represents the resolution for the HCO$^+$ line.
\label{g333_redblue}
}
\end{center}
\end{figure}

Figure \ref{g333_redblue} shows the red and blue HCO$^+$ components of G333.234--00.061. The redshifted and blueshifted velocity wings of MM2 spatially overlap, indicating a bipolar outflow closely oriented along the line of sight. A filament-like structure connects this outflow feature through MM1 to an HCO$^+$ source toward the southeast.

\subsubsection{Masers}\label{sec:masers}
Masers are seen throughout G333.234--00.061, as seen in Figure \ref{g333_cont}. Here, we briefly discuss some background about these maser types, and then we infer the evolutionary state and some properties of MM1 and MM2.

Four masers are commonly associated with high-mass star-forming regions: two types of methanol (CH$_3$OH) masers, water masers, and hydroxyl (OH) masers. Methanol masers are categorized into two different classes. Class I methanol masers are collisionally pumped \citep[e.g.,][]{Leurini2004} while class II methanol masers are radiatively pumped \citep{Cragg1992}. Class I masers are often offset from the nearest continuum peak \citep[e.g.,][]{Kurtz2004} and found in shocked regions, particularly those associated with molecular outflows, \citep[e.g.,][]{Voronkov2006,Cyganowski2009,Gan2013}. Although typically found within high-mass star-forming regions, class I masers have also been reported toward supernova remnants \citep[e.g.,][]{Pihlstrom2014}. Class II methanol masers have thus far $only$ been found associated with high-mass star-forming regions \citep[e.g.,][]{Xu2008}. Class I masers are suggested to be associated with star-forming regions at an earlier evolutionary stage (e.g., infrared dark clouds) than class II masers \citep[e.g.,][]{Ellingsen2006}. Class I masers also can occur after class II masers due to, e.g., expanding \ion{H}{2} regions \citep{Voronkov2010}, but MM1 and MM2 appear to pre-date these evolutionary stages.

Like class I methanol masers, water masers are also collisionally pumped and associated with shocks and outflows \cite[e.g.,][]{Elitzur1989}, and like class II methanol masers, OH masers are radiatively pumped \citep{Cragg2002}. Water masers are suggested to appear before OH masers \citep{Forster1989}, but after methanol masers \citep{Breen2007}.

In G333.234-00.061 all four masers (methanol class I and class II, water, and OH) are detected. Their locations are indicated in Figure \ref{g333_cont}. Both cores are associated with a class II methanol maser \citep[6.7~GHz,][]{Caswell2011}, while only MM1 contains a class I methanol maser centered on the source \citep[95.1~GHz,][]{Ellingsen2005}. Additional class I methanol masers at 36 and 44 GHz \citep{Voronkov2014} are scattered throughout the region rather than being centered on a particular source, which may indicate active outflows. Only MM2 is associated with water \citep[22~GHz,][]{Breen2010} and hydroxyl \citep[1.7~GHz,][]{Caswell1998} masers. Although many studies \citep[e.g.,][]{Forster1989,Ellingsen2006,Breen2007} suggest that the appearance of masers from earlier to later evolutionary stages (often with overlap) follows the sequence ``Class I Methanol$\rightarrow$Class II Methanol$\rightarrow$Water$\rightarrow$OH", exceptions are frequently seen \citep[e.g.,][]{Voronkov2006,Cyganowski2012}. Such inconsistencies are likely due to the complexity of high-mass star formation (e.g., large clusters of stars with a variety of masses contained within diverse environments). Nevertheless, if masers generally follow this evolutionary trend, MM2 is likely more evolved due to its lack of a class I methanol maser and MM1's lack of water and OH masers. 

Although class I masers are typically associated with outflows, MM2's strong outflow does not appear to be associated with a class I maser. The outflow may have broken through most of the circumstellar material and is no longer producing collisions capable of exciting the class I maser. MM1, on the other hand, could have an outflow that has yet to break through its natal environment, and thus produces shocked gas capable of exciting a class I methanol maser.

As discussed in Section \ref{sec:g333_lines}, from $^{13}$CS(2--1) we derive a core velocity of $-84.7$~km$\,$s$^{-1}$ for MM1, and two core velocities of $-86.0$~km$\,$s$^{-1}$ and $-89.8$~km$\,$s$^{-1}$ (with the latter the brightest) for MM2. For MM1, the 95.1~GHz class I methanol maser has a velocity of $-87.4$~km$\,$s$^{-1}$ \citep{Ellingsen2005} and the class II methanol maser has a velocity of --85.3~km$\,$s$^{-1}$ \citep{Caswell2011}. For MM2, the class II methanol, water, and OH masers have velocities of --91.9~km$\,$s$^{-1}$, --88~km$\,$s$^{-1}$, and --84~km$\,$s$^{-1}$ respectively \citep{Caswell2011,Breen2010,Caswell1998}. As expected, the radiatively pumped (class II methanol) maser for MM1 is closer to the velocity of the core than the collisionally pumped (class I methanol) maser, strengthening the hypothesis that the class I methanol maser for MM1 is due to an outflow. MM2 is more complex, perhaps due to its two velocity components. The class II methanol and water masers are closer to the higher velocity component, while the OH maser is closer to the lower velocity component. The difference between the velocities of the two radiatively pumped (class II and OH) masers may indicate multiple cores in MM2.




\subsubsection{Two high-mass cores at different evolutionary stages}
From the preceding discussion, MM2 is likely at a later evolutionary stage than MM1 for the following reasons: \\
(1) MM2 has $Spitzer$ IRAC emission while MM1 does not, suggesting much warmer temperatures for MM2.\\
(2) MM2 has a large outflow, while MM1 does not have a definitive outflow.\\
(3) $^{13}$CS has stronger emission on MM2 even though MM1 is brighter in the continuum, suggesting MM1 could be colder.\\
(4) The detected maser species for MM1 and MM2 indicate that MM2 is likely at a later evolutionary stage.

These cores are most likely protostellar because they seem to lack free-free emission (see Section \ref{sec:g333_cont}), have masers that are expected to be absent in prestellar cores, and the fact (for MM2) that the compact mid-IR colors indicate a deeply embedded source. The ratio of the integrated flux densities of MM1 and MM2 is 1.25; since MM1 is likely colder, it should be at least 25\% more massive than MM2.  The projected separation between MM1 and MM2 is only 0.12~pc; therefore two of the most massive protostellar cores known are forming in close proximity at two different evolutionary stages, with the more massive core at an earlier evolutionary stage.



During the high-mass star formation process, more massive cores are thought to both collapse and accrete faster and thus evolve faster \citep[e.g.,][]{Zinnecker2007}. Therefore, it is surprising that two high-mass cores can form within close proximity with the younger core more massive than the older core. We offer the following possible explanations for these observations:\\
\noindent(1) MM2 may have started to form at a much earlier time than MM1. This would suggest that multiple epochs of massive star formation can happen in close proximity.\\
(2) In the past, MM2 may have had a higher accretion rate than MM1, and thus evolved more quickly. However, MM1's accretion rate may now be higher, but it has yet to become as evolved as MM2.\\
(3)  The relationship between free-fall time, $t_{ff}$, and density, $\rho$, is $t_{ff} \propto \rho^{-1/2}$, and this timescale likely indicates the approximate collapse time for forming cores MM1 and MM2. MM2's initial environment may have had a higher density and thus collapsed and evolved quicker. However, MM1's initial gas reservoir may have been more massive and has recently been accreted to make the core more massive than MM2.\\
(4) The apparently more massive core MM1 might consist of two or more unresolved lower mass cores.  Its earlier evolutionary state would then be easily explained since lower mass cores should evolve more slowly.  Indeed, for intermediate mass cores in the GGD27 complex, with higher angular resolution observations, \citet{FL2011} resolved an apparently more massive core into two lower mass cores separated by only 1700~AU, well below our resolution of $\sim$10000~AU.\\
(5) MM2 could actually be more massive. MM1 could have a higher temperature than MM2 causing underestimates of the gas mass, but we believe this as unlikely given the discussion in this section. Moreover, the cores could have different dust opacities ($\kappa_\nu$) due to different chemistry or densities.\\
(6) For core accretion models, radiative and mechanical feedback mechanisms cause the star's final mass to be less than the prestellar core \citep[e.g.,][]{Tan2014}. Such feedback may have been acting longer on MM2 than MM1, causing MM2 to be less massive than MM1 even though MM2 was initially a more massive prestellar core. The possibility is unlikely since the evolutionary difference between the two cores is not drastically different (both are likely protostellar).




\subsection{G345.144--00.216}\label{sec:G345}

\subsubsection{Continuum}\label{sec:g345_cont}
\begin{figure}[ht!] Ê
\begin{center}
\includegraphics[scale=0.3]{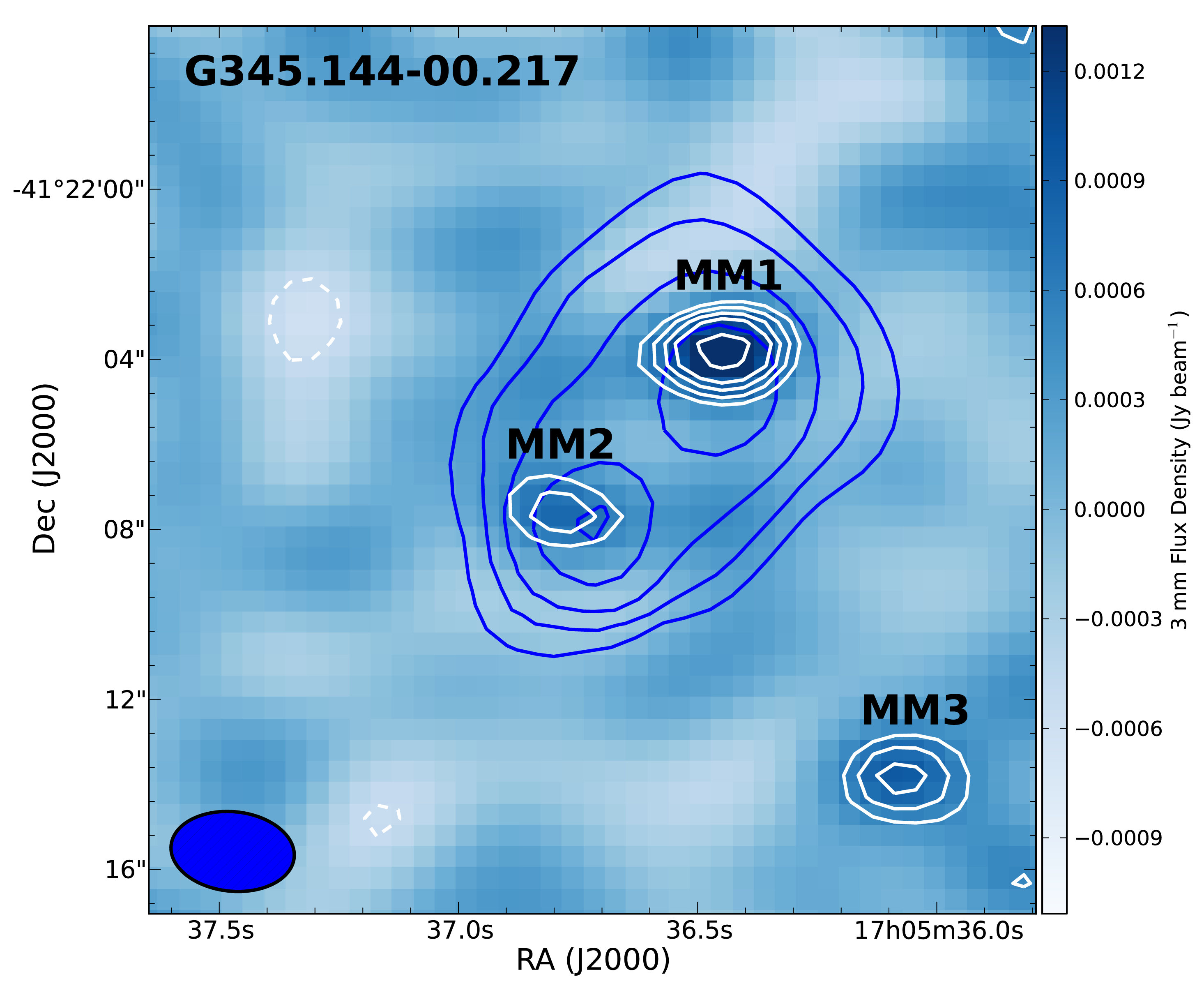}
\caption{The ATCA 3.3~mm continuum image for G345.114--00.216, corrected for the primary beam. White contours and color-scale are shown for the 3.3~mm continuum, with contours shown for [--3, 3, 4, 5, 6, 8]~$\times$~$\sigma_{3.3\rm{mm}}$, where $\sigma_{3.3\rm{mm}}=0.17$~mJy$\,$beam$^{-1}$ . Blue contours are the $Spitzer$ IRAC 8.0~$\mu$m with contour levels of [200, 300, 500, 1000, 1800]~$\times$~1 MJy$\,$sr$^{-1}$. There are no known masers in the clump. \label{g345_cont}
}
\end{center}
\end{figure}

The 3.3~mm continuum map for G345.144--00.216 (Figure \ref{g345_cont}) reveals three unresolved cores ($<$2$\farcs$3 or 0.02 pc). The brightest core and the dimmest core (henceforth MM1 and MM2 respectively) are both coincident with $Spitzer$ IRAC and MIPS 24~$\mu$m emission (8.0 $\mu$m is shown), with the brightest IRAC emission centered on MM2. The southwest core, MM3, is devoid of any IRAC or 24~$\mu$m MIPS emission. The peak fluxes for MM1, MM2, and MM3 are 1.5, 0.80, and 0.94~mJy$\,$beam$^{-1}$, respectively, and these fluxes were used to calculate the masses. We assume the same values for $\kappa_{3.3\,\rm{mm}}$ and the gas-to-dust mass ratio  (0.22~cm$^2$~g$^{-1}$ and 100 respectively) and reproduce the same error analysis as Section \ref{sec:g333_cont}. For cores with temperatures of 100~K, we find the masses to be 0.45$^{+0.61}_{-0.30}$, 0.23$^{+0.32}_{-0.15}$, and 0.28$^{+0.37}_{-0.18}$~$M_\sun$ for MM1, MM2, and MM3 respectively. Assuming a colder temperature of 50~K, the core masses approximately double to 0.92$^{+1.07}_{-0.73}$ 0.48$^{+0.55}_{-0.38}$, and 0.56$^{+0.65}_{-0.45}$~$M_\sun$ for MM1, MM2, and MM3 respectively. MM3's mass may be much higher since it lacks IRAC emission, suggesting it is colder than the other two cores. MM1 and MM2 are also seen with MIPSGAL at 24~$\mu$m while MM3 is undetected.



The MIRIAD task \texttt{maxfit} was used to locate the peak intensities, placing MM1 at R.A. (J2000.0$)=17^{\rm{h}}05^{\rm{m}}36.447\fs29$ and decl.~$=-41\arcdeg22\arcmin03\farcs8$, MM2 at R.A. (J2000.0$)=17^{\rm{h}}05^{\rm{m}}36\fs78$ and decl.~$=-41\arcdeg22\arcmin07\farcs6$, and MM3 at R.A. (J2000.0$)=17^{\rm{h}}05^{\rm{m}}36\fs08$ and decl.~$=-41\arcdeg22\arcmin13\farcs8$. No known masers are found in the clump.


\subsubsection{Molecular lines}\label{sec:g345_lines}
\begin{figure*} [ht!]Ê
\begin{center}
\includegraphics[scale=0.7]{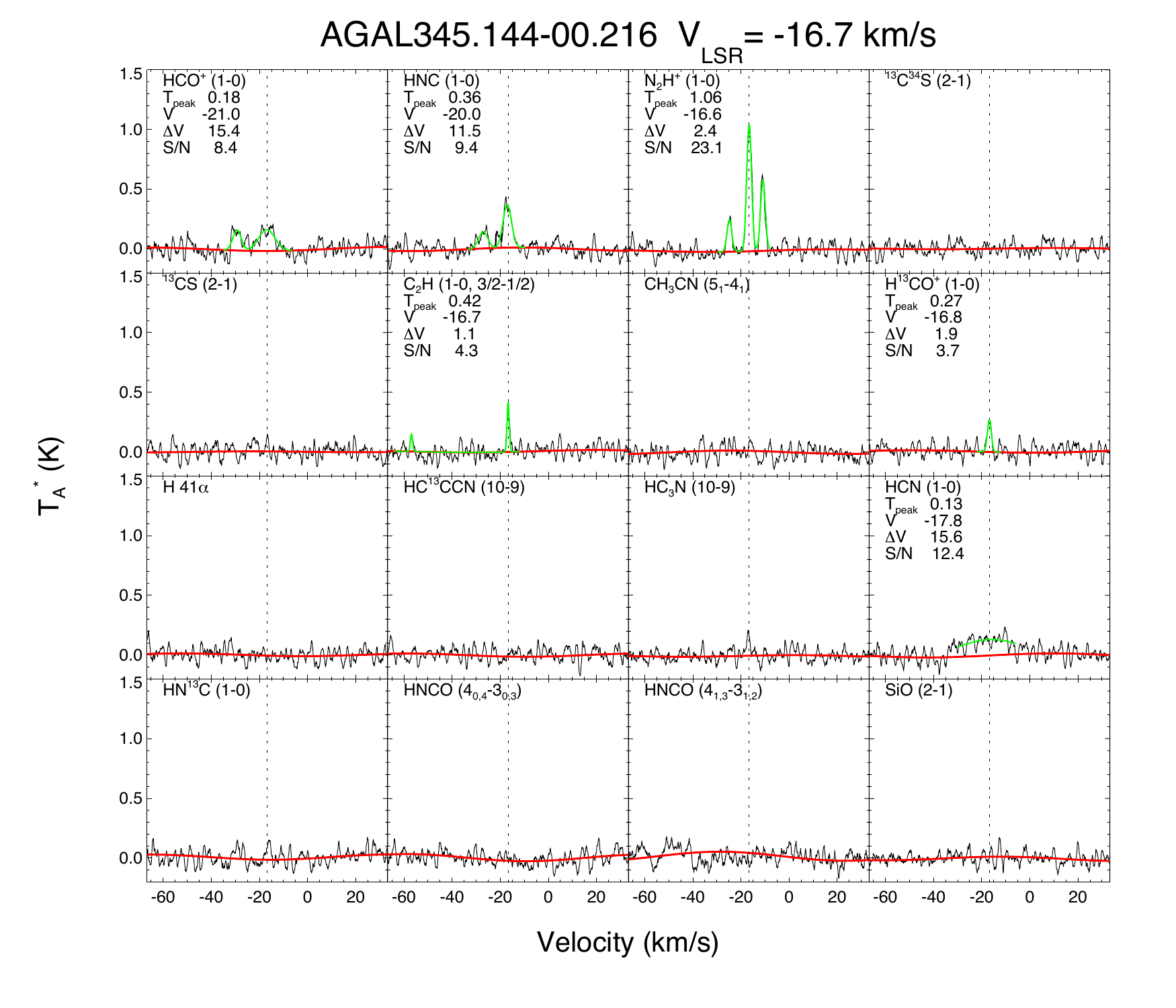}
\caption{MALT90 (resolution of 38$\arcsec$) observations and fit results of G345.144--00.216 for 16 observed lines (J. Rathborne et al., in preparation). Caption is the same as Figure \ref{g333_malt90}. Two line fits are shown for HCO$^+$(1--0) and HNC(1--0), but only the fit parameters of the strongest line is shown. \label{g345_malt90}
}
\end{center}
\end{figure*}

Figure \ref{g345_malt90} shows the 16 lines from MALT90, with the N$_2$H$^+$ line the brightest. The HNC spectrum has two obvious peaks that may represent two velocity components or a self-absorption feature. The HCO$^+$ spectrum also hints at two peaks with about the same velocities as HNC. The $^{13}$CS (1--0), C$_2$H (1--0), H$^{13}$CO$^+$ (1--0), HC$_3$N (1--0), HCN (1--0), and HN$^{13}$C (1--0) lines appear to be detected or marginally detected.


\begin{figure} [ht!]Ê
\begin{center}
\includegraphics[scale=0.063]{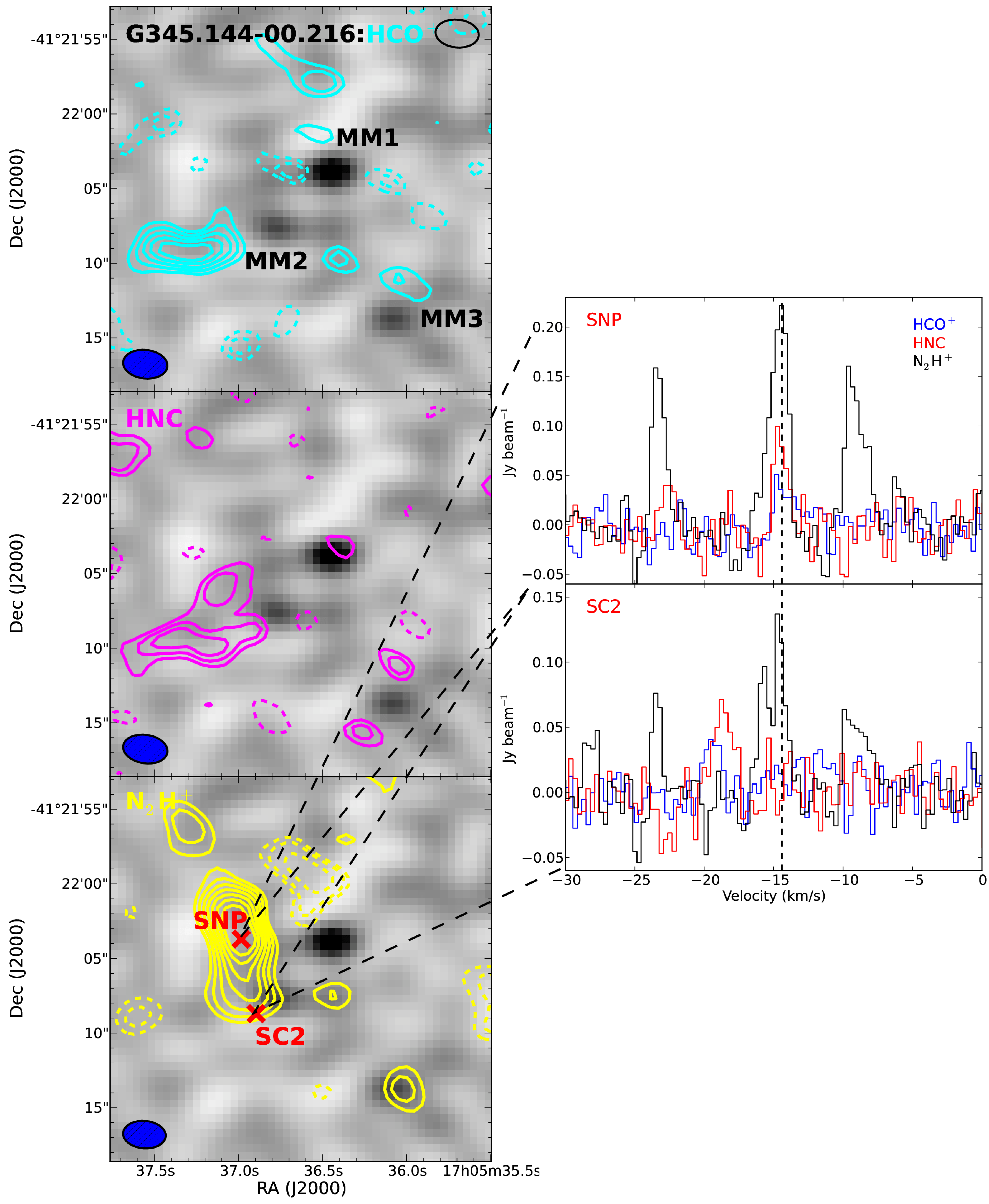}
\caption{G345.144--00.216 ATCA integrated intensity maps of the molecular lines HCO$^+$(1--0), HNC(1--0), and N$_2$H$^+$(1--0) on top of a grayscale map of the 3.3~mm continuum. Reference contours are shown for [--5, --4, --3, 3, 4, 5, 6, 7, 8, 9, 12, 15]~$\times~f$ for $f_{\rm{HCO}^+} =  0.025$, $f_{\rm{HNC}} =  0.025$, and $f_{\rm{N}_2\rm{H}^+} =  0.06$~Jy$\,$beam$^{-1}$~km$\,$s$^{-1}$. The open black beam (top panel, top right) represents the continuum resolution while the filled blue beam (bottom left) represents the panel's molecular line resolution. Lines were integrated based on the following velocity ranges: HCO$^+$: --22.1 to --11.9~km$\,$s$^{-1}$, HNC: --22.1 to --11.9~km$\,$s$^{-1}$, and \nnhp: --26.0 to --7.1~km$\,$s$^{-1}$. Selected spectra (Gaussian smoothed) are shown on the right for the positions shown in the N$_2$H$^+$ map, with a reference dashed line drawn at --14.4~km$\,$s$^{-1}$.
\label{g345_lines}
}
\end{center}
\end{figure}

The G345.144--00.216 ATCA integrated intensity maps are shown in Figure \ref{g345_lines}. HCO$^+$, HNC, and N$_2$H$^+$ were detected while $^{13}$CS was undetected. The three compact cores unexpectedly lack strong line emission. Self-absorption may suppress the ground state (1--0) line intensity below detectable levels; such deep self-absorption has been seen toward other high-mass clumps, e.g., in the HCO$^+$ and HNC (1--0) lines toward the infrared dark cloud G028.23--00.19 \citep{Sanhueza2013}. The self-absorption in G345.144--00.216 could be similar to what we see in G333.234--00.061, except that G333.234--00.061 has a strong outflow delineated by optically thinner HCO$^+$ emission.

\begin{figure} [ht!]Ê
\begin{center}
\includegraphics[scale=0.3]{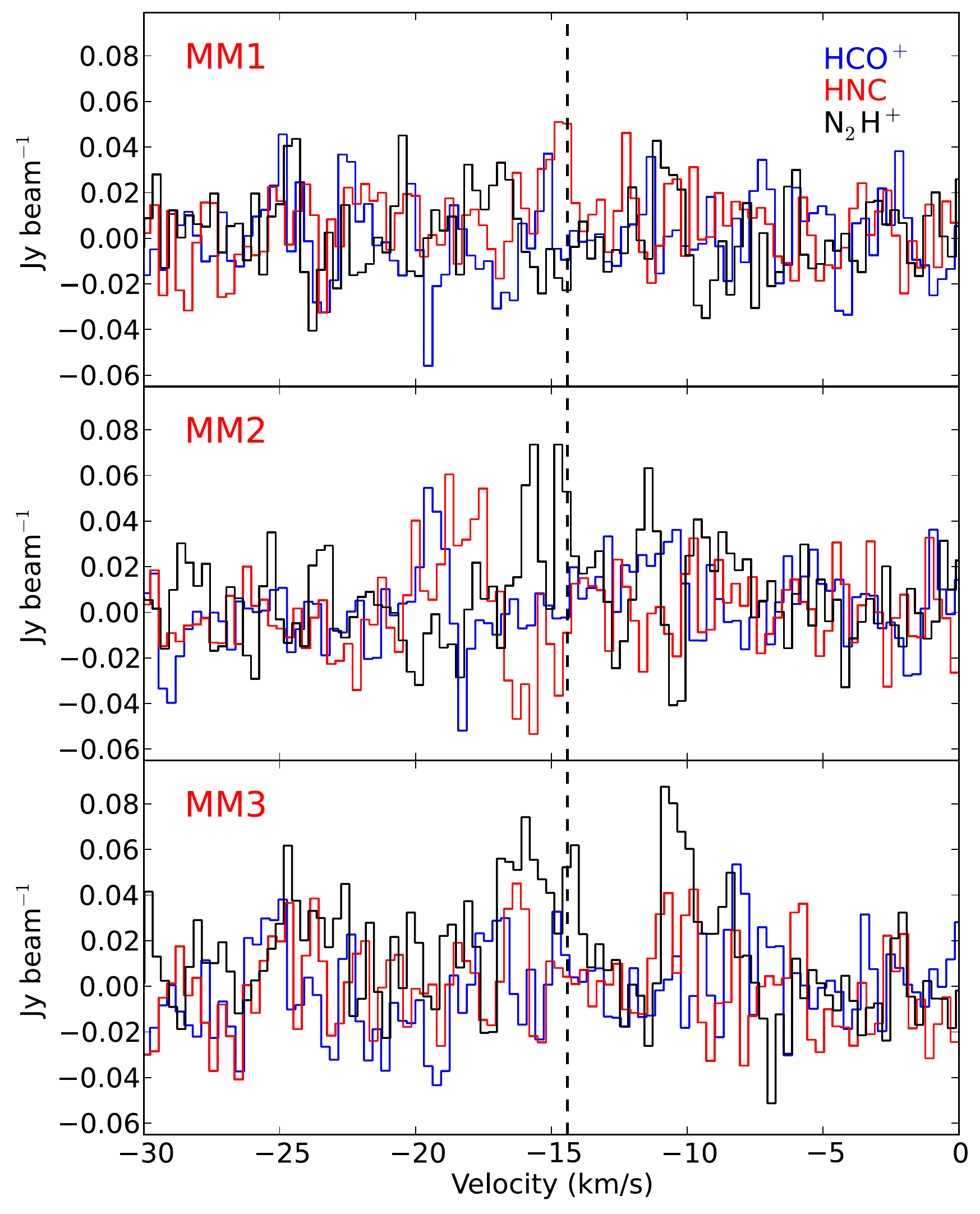}
\caption{G345.144--00.216 ATCA spectra (Gaussian smoothed) are shown for the peak intensity pixels of MM1, MM2, and MM3, with a reference dashed line drawn at --14.4~km$\,$s$^{-1}$.
\label{g345_spectra_peaks}
}
\end{center}
\end{figure}

To explore the possibility of self-absorption, we analyzed spectra throughout the clump. On the right of Figure \ref{g345_lines}, two example spectra are shown: one at the N$_2$H$^+$ emission peak (SNP, Spectra at \nthp\ Peak) and one near MM2 (SC2, Spectra at Core 2). Toward the SNP position, the lines are detected at this position at the same velocity of approximately --15~km$\,$s$^{-1}$.  The spectra for SC2 show that near MM2, there are two apparent velocity components in the \nthp\ spectrum whose average velocity is approximately --15~km$\,$s$^{-1}$. The HCO$^+$ and HNC line profiles, on the other hand, are offset in velocity from the N$_2$H$^+$ velocity peak and are also slightly offset from each other. If self-absorption is not occurring, there must be four distinct velocity components at this location (two indicated by N$_2$H$^+$ and one each indicated by HCO$^+$ and HNC). However, it is extremely unlikely that these three molecules each trace a different velocity component, especially since they all trace one component at about --15~km$\,$s$^{-1}$ at the SNP \nthp\ peak. Therefore, these lines are likely self-absorbed with a central velocity at approximately --15~km$\,$s$^{-1}$. Only the blue components for HCO$^+$ and HNC survive, with HCO$^+$ appearing to be slightly more self-absorbed. Moreover, spectra centered on the intensity peak of all three cores (Figure \ref{g345_spectra_peaks}) show almost no HCO$^+$ or HNC emission (with MM2 having the only detection at $\sim$18~\kms) and weak N$_2$H$^+$ emission for MM2 and MM3. These results are consistent with \citet{Sanhueza2012}, which found that clump optical depths are typically slightly higher for HCO$^+$ than HNC, while the optical depths for N$_2$H$^+$ are over an order of magnitude smaller.

%
%
%

G333.234--00.061's self-absorption was much more obvious since its strong outflow shows broad velocity wings in the spectra that were not self-absorbed. No outflow was detected in G345.144--00.216, resulting in the non-detection of most of the observed lines toward the dust cores. These results lead to an interesting conclusion: if no lines are detected toward a core, it does not necessarily mean the molecule is absent. Therefore, observations of less abundant (e.g., isotopologues) or more highly excited transitions are needed to accurately trace these extremely high column density environments.

\section{N$_2$H$^+$ Poor Sources}\label{sec:n2hppoor}
\subsection{G351.409+00.567 and G353.229+00.672}\label{sec:G351G353}
The ATCA detected no 3.3~mm continuum sources toward G351.409+00.567 and G353.229+00.672, and the literature reports no masers. The 1$\sigma$ 3.3~mm continuum sensitivity for these two sources are 0.12 and 0.16~mJy$\,$beam$^{-1}$ respectively. If we assume undetected cores are no more than four times brighter than the rms noise within a given beam size, $\kappa_{3.3\,\rm{mm}} = 0.22$~cm$^2$~g$^{-1}$, a gas-to-dust mass ratio of 100, and a cold core dust temperature of $T_D = 10$~K, we conclude that it is likely that there are no cores smaller than the beam more massive than 1.4~$M_\sun$ and 1.9~$M_\sun$ in G351.409+00.567 and G353.229+00.672 respectively. However, it is possible there are cores that are more extended than the synthesized beam but have extended emission below the sensitivity limit.

\begin{figure*} [ht!]Ê
\begin{center}
\includegraphics[scale=0.7]{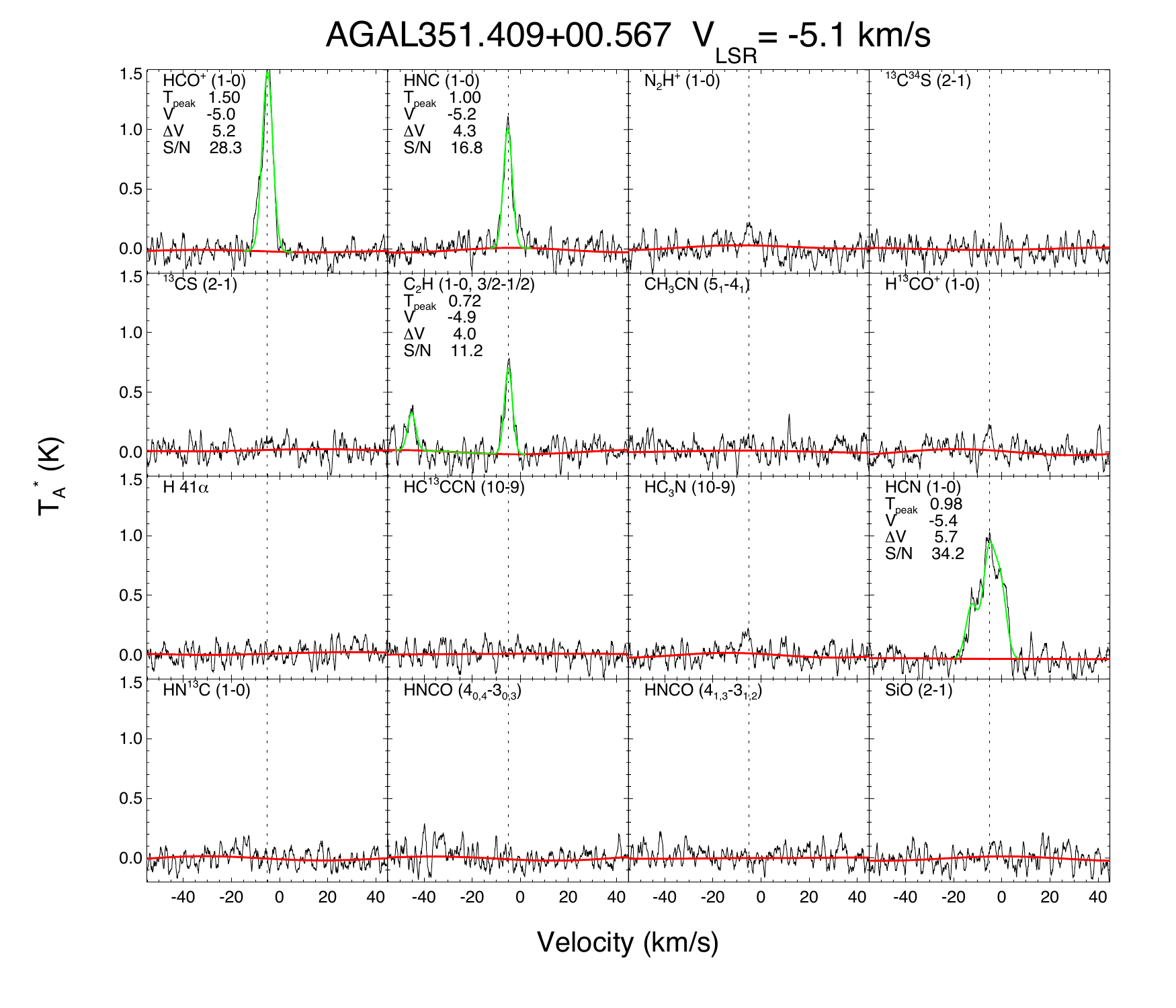}
\caption{MALT90 (resolution of 38$\arcsec$) observations and fit results of G351.409+00.567 for 16 observed lines (J. Rathborne et al., in preparation). Caption is the same as Figure \ref{g333_malt90}. \label{g351_malt90}
}
\end{center}
\end{figure*}

\begin{figure*} [ht!]Ê
\begin{center}
\includegraphics[scale=0.7]{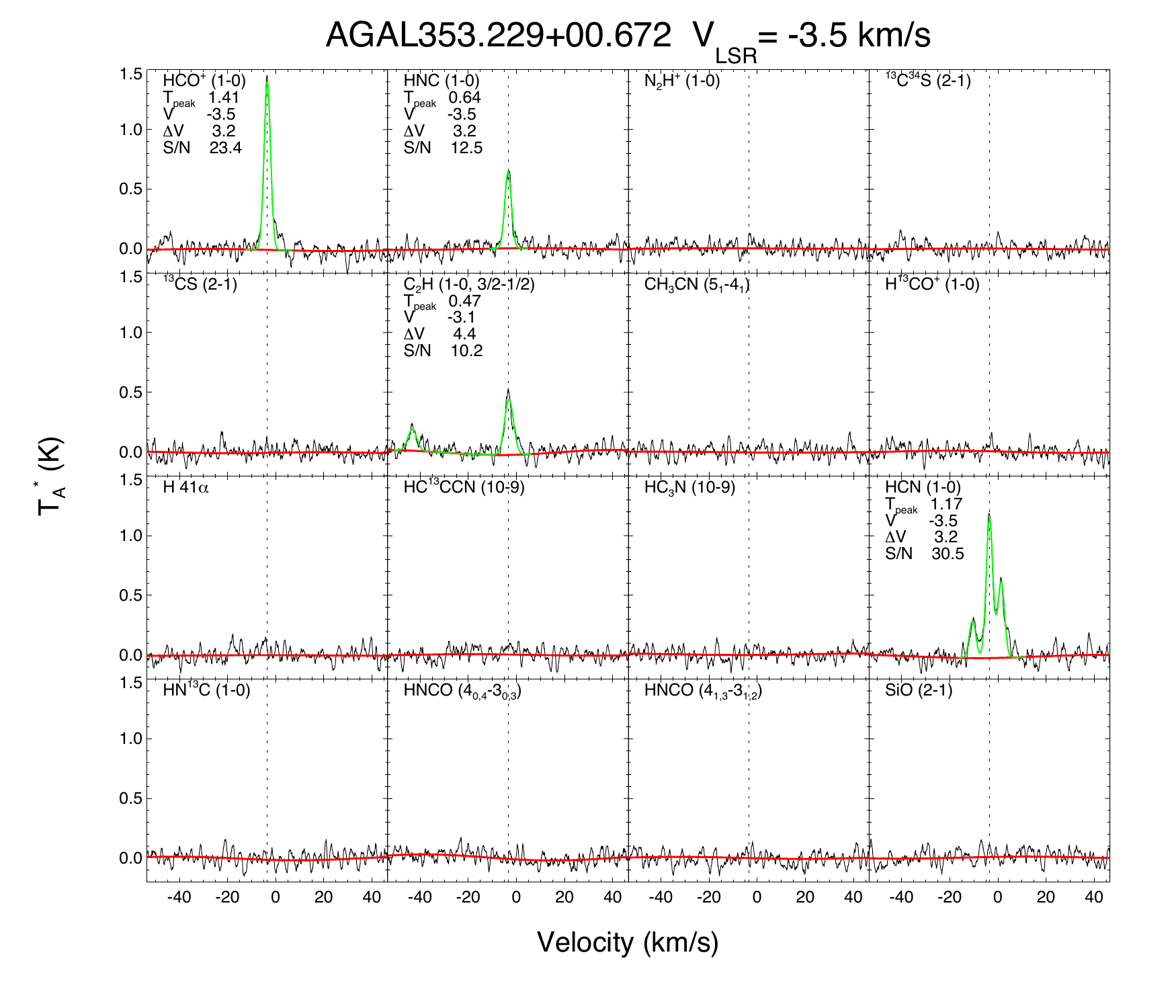}
\caption{MALT90 (resolution of 38$\arcsec$) observations and fit results of G353.229+00.672 for 16 observed lines (J. Rathborne et al., in preparation). Caption is the same as Figure \ref{g333_malt90}. \label{g353_malt90}
}
\end{center}
\end{figure*}

Molecular lines from MALT90 are shown for G351.409+00.567 and G353.229+00.672 in Figures \ref{g351_malt90} and \ref{g353_malt90} respectively. G351.409+00.567 and G353.229+00.672 have integrated intensity ratios [I(N$_2$H$^+$)/I(HCO$^+$)] of 0.07 and 0.09 respectively. Each source has strong C$_2$H emission which is commonly considered a PDR tracer \citep[e.g.,][]{Jansen1995}.

\begin{figure} [ht!]Ê
\begin{center}
\includegraphics[scale=0.075]{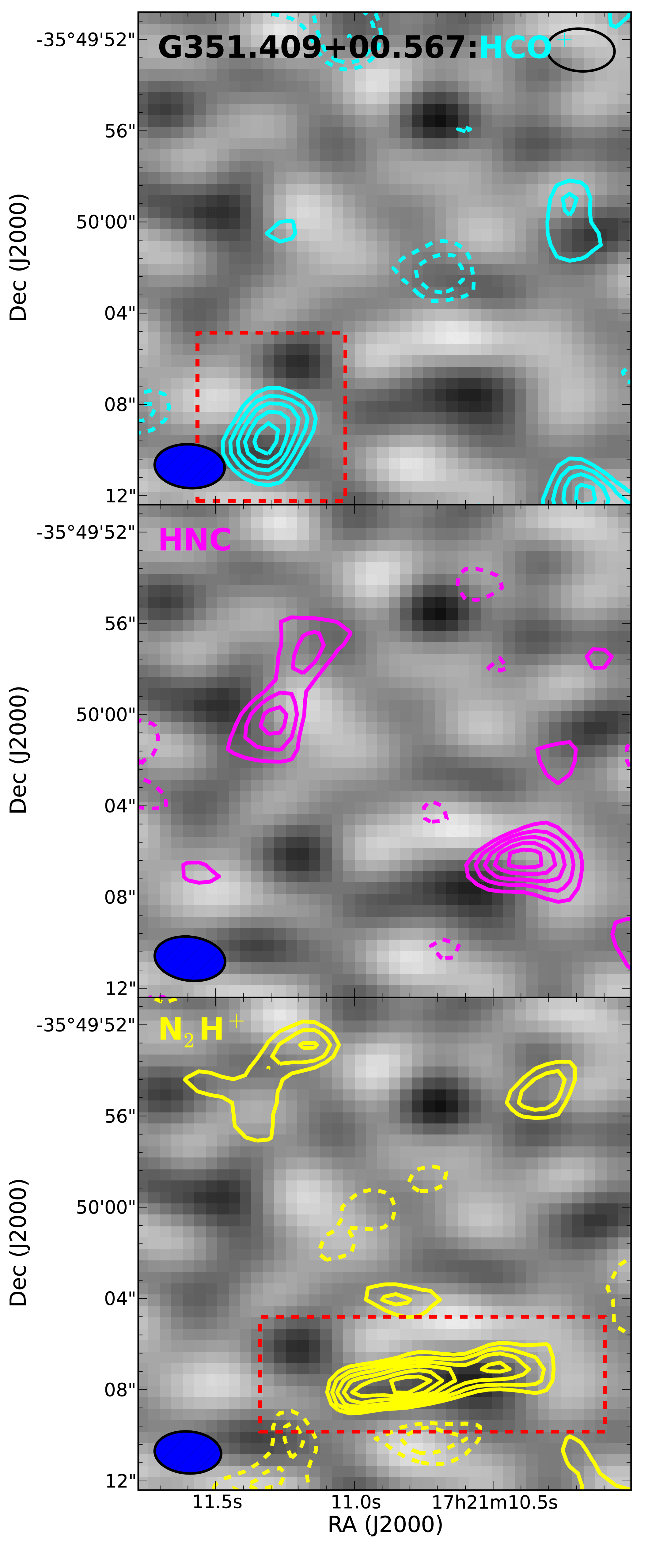}
\caption{G351.409+00.567 ATCA observations. Caption is the same as Figure \ref{g345_lines}, with $f_{\rm{HCO}^+} =  0.0185$, $f_{\rm{HNC}} =  0.0185$, and $f_{\rm{N}_2\rm{H}^+} =  0.02$~Jy$\,$beam$^{-1}$~km$\,$s$^{-1}$. No sources are found above 4$\sigma$ in the continuum. Lines were integrated based on the following velocity ranges: HCO$^+$: --7.6 to --1.0~km$\,$s$^{-1}$, HNC: --7.6 to --1.9~km$\,$s$^{-1}$, and \nnhp: --13.6 to --1.91~km$\,$s$^{-1}$. The red-dashed boxes show locations where the positive pixels were integrated to investigate the small-scale [I(N$_2$H$^+$)/I(HCO$^+$)] ratio.
\label{g351_lines}
}
\end{center}
\end{figure}

\begin{figure} [ht!]Ê
\begin{center}
\includegraphics[scale=0.075]{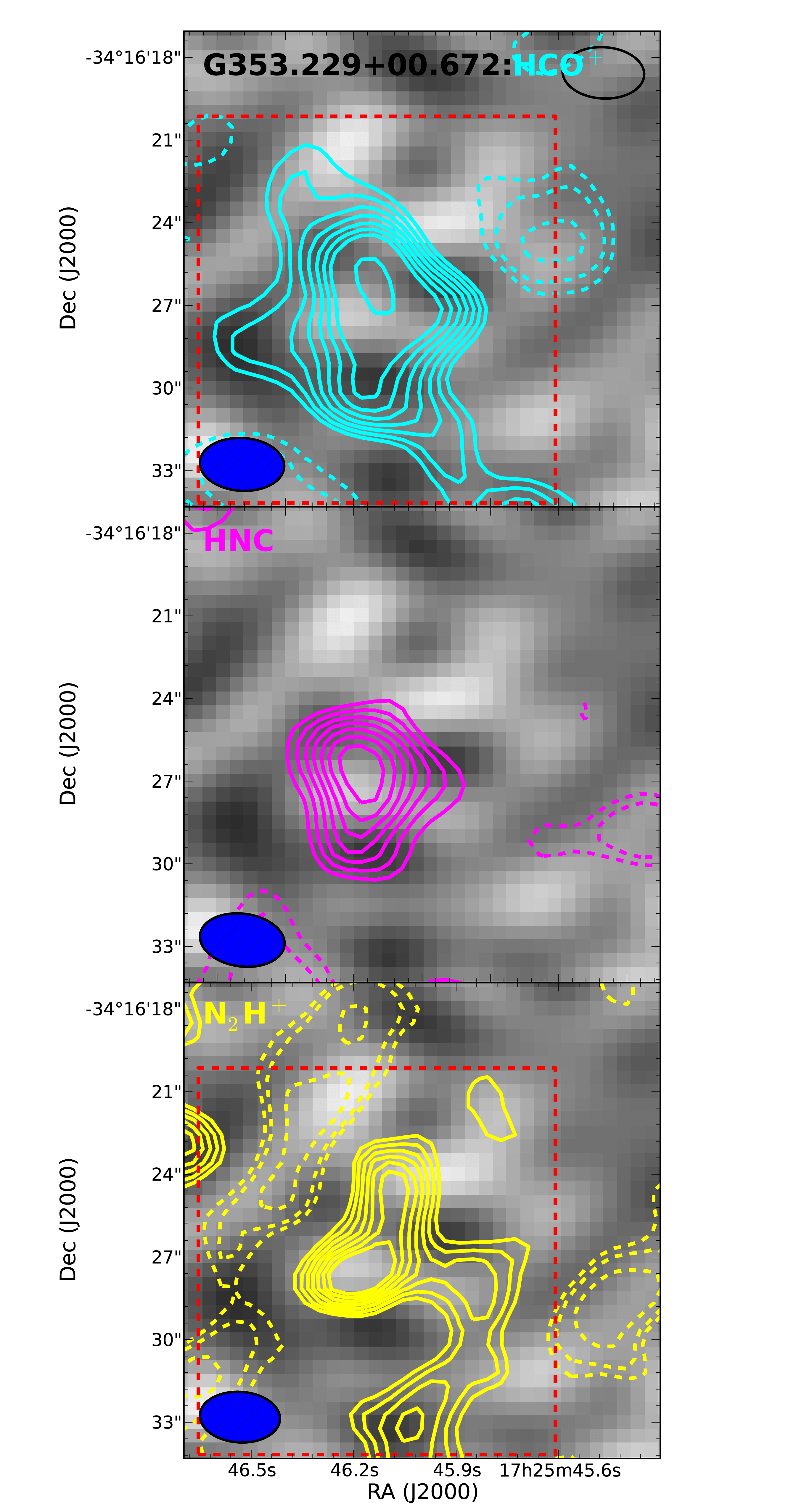}
\caption{G353.229+00.672 ATCA observations. Caption is the same as Figure \ref{g345_lines}, with $f_{\rm{HCO}^+} =  0.035$, $f_{\rm{HNC}} =  0.035$, and $f_{\rm{N}_2\rm{H}^+} =  0.022$~Jy$\,$beam$^{-1}$~km$\,$s$^{-1}$.  No sources are found above 4$\sigma$ in the continuum. Lines were integrated based on the following velocity ranges: HCO$^+$: --5.8 to 2.5~km$\,$s$^{-1}$, HNC: --6.6 to 2.5~km$\,$s$^{-1}$, and \nnhp: --13.1 to 5.4~km$\,$s$^{-1}$. The red-dashed boxes show locations where the positive pixels were integrated to investigate the small-scale [I(N$_2$H$^+$)/I(HCO$^+$)] ratio. \label{g353_lines}
}
\end{center}
\end{figure}

Figures \ref{g351_lines} and \ref{g353_lines} show integrated intensity maps of the three lines (HCO$^+$, HNC, and N$_2$H$^+$) detected toward both clumps. In G353.229+00.672, the HNC, N$_2$H$^+$, and HCO$^+$ line emission is spatially coincident, but in G351.409+00.567, each line arises from a different location. The lack of spatial coincidence of the lines for G351.409+00.567 could be artifacts due to the interferometer being insensitive to large-scale structure.

To measure the integrated fluxes of the HCO$^+$ and N$_2$H$^+$ lines throughout the moment maps, we masked the negative pixels and summed the positive pixels based on the red-dashed regions outlined in Figures \ref{g351_lines} and \ref{g353_lines}. Since line fluxes are not coincident for G351.409+00.567, the integrated fluxes were drawn from different areas of the map where the line fluxes are concentrated. For G351.409+00.567 we find integrated flux densities of 0.31 Jy~km$\,$s$^{-1}$ and 0.65 Jy~km$\,$s$^{-1}$ for HCO$^+$ and N$_2$H$^+$ respectively. For G353.229+00.672 we find integrated flux densities of 3.7 and 2.0 Jy~km$\,$s$^{-1}$  for HCO$^+$ and N$_2$H$^+$ respectively. On these small angular scales, the [I(N$_2$H$^+$)/I(HCO$^+$)] ratios are about 2 and 0.5 for G351.409+00.567 and G353.229+00.672 respectively, which is much less anomalous than the large-scale MALT90 line observations (0.04 and 0.12 respectively). If these line fluxes are diluted to the MALT90 beam (38$\arcsec$), the fluxes would all be less than 0.4~K~km$\,$s$^{-1}$, and thus all lines would either be marginally detected or undetected. Since MALT90 reports significant detections for HCO$^+$ (and HNC) for G351.409+00.567 and G353.229+00.672 ($8.3\pm0.3$ and $4.8\pm0.2$~K~km$\,$s$^{-1}$ respectively, J. Rathborne et al. in preparation), the majority of the HCO$^+$ line emission of these sources and the presence of the N$_2$H$^+$ poor anomaly occurs at large scales rather than small scales. 





\subsection{MALT90 and N$_2$H$^+$ Poor Sources}\label{sec:MALT90n2hppoor}
Since HCO$^+$ is usually much more optically thick than N$_2$H$^+$, self-absorption cannot explain N$_2$H$^+$ poor sources, as it does the N$_2$H$^+$ rich sources. The ATCA data suggest that sources only appear N$_2$H$^+$ poor on large angular scales. Thus, to properly investigate the N$_2$H$^+$ poor sources, it is best to use large-scale (MALT90) observations for interpretation. 

We create an \nthp\ poor sample based on the MALT90 catalog (J. Rathborne et al., in preparation) with the following criteria:

\noindent(1) Catalog sources are only kept if they either have (a) no detection in \nthp\ but a detection in HCO$^+$, or (b) if they have integrated intensity ratios: I(N$_2$H$^+$)/I(HCO$^+$)] + $\sigma_r$~$< 0.2$, where $\sigma_r$ is the error of the integrated intensity ratio. These cutoffs reduce our sample to about 20\% of the total MALT90 sources. \\
(2) For sources that had no detection in \nthp\ but a detection in HCO$^+$, we only consider sources with I(HCO$^+$)~$>$~2.5 K~km$\,$s$^{-1}$. In order to keep sources with I(N$_2$H$^+$)/I(HCO$^+$)]~$\lesssim 0.2$, I(HCO$^+$)~$<$~2.5~K~km$\,$s$^{-1}$ constrains I(N$_2$H$^+$) values to be less than 0.5~K~km$\,$s$^{-1}$. 0.5~K~km$\,$s$^{-1}$ is just above the typical MALT90 integrated intensity detection limit, and thus this criteria selects sources with I(N$_2$H$^+$)/I(HCO$^+$)] $\lesssim 0.2$. \\
(3) In order to only retain the clumps with the most reliable I(HCO$^+$) measurements, sources were removed with an I(HCO$^+$) signal-to-noise ratio $<$ 6.\\
(4) Sources were removed  that lie within a Galactic longitude of $-2^\circ < l < 4^\circ$. These sources lie within the central molecular zone of the Galaxy where clumps are subject to different conditions and chemical processes than those in the rest of Galaxy.\\
(5) Sources were removed when the fine structure lines of N$_2$H$^+$ fall out of the frequency range of the MALT90 bandpass. The velocity cutoff depends on the sky frequency; sources were typically rejected when their systemic velocity was less than --93~km$\,$s$^{-1}$.\\
(6) Sources were removed that have HCO$^+$ linewidths $>15$~\kms. These sources comprised of $\sim$10\% of the remaining sample and are therefore anomalous themselves. Such sources may poorly represent the \nthp\ poor sample here.\\
(7) The remaining 127 sources were visually inspected, and three additional sources (AGAL010.356-00.149\_B, AGAL332.191-00.047\_S, and AGAL354.838+00.372\_S) were removed due to low signal to noise that caused significant difficulty in providing good fits to the \nthp\ spectra.

The evolutionary classifications (as discussed in Section \ref{sec:classification}) for the 124 remaining N$_2$H$^+$ poor clumps consisted of 41 PDRs, 36 \ion{H}{2} regions, 5 Protostellar, 8 Quiescent, and 34 Unknown sources. Of the total 3566 clumps MALT90 observed, there were 362, 938, 796, 689, and 781 sources classified as PDRs, \ion{H}{2} regions, Protostellar, Quiescent, and Unknown, respectively. The probability that a random MALT90 clump is an N$_2$H$^+$ poor source based on our criteria is 11.3\%, 3.8\%, 0.6\%, 1.1\%, and 4.4\% for PDRs, \ion{H}{2} regions, Protostellar, Quiescent, and Unknown respectively. Immediately it is evident that these N$_2$H$^+$ poor clumps are primarily associated with regions with high ionization (i.e., PDRs and \ion{H}{2} regions, with the former very evident) as well as the clumps with Unknown classification. The Unknown clumps come from diverse environments, so we cannot draw any general scientific conclusions about the chemistry from these clumps.

\begin{figure*} [ht!]Ê
\begin{center}
\includegraphics[scale=0.3]{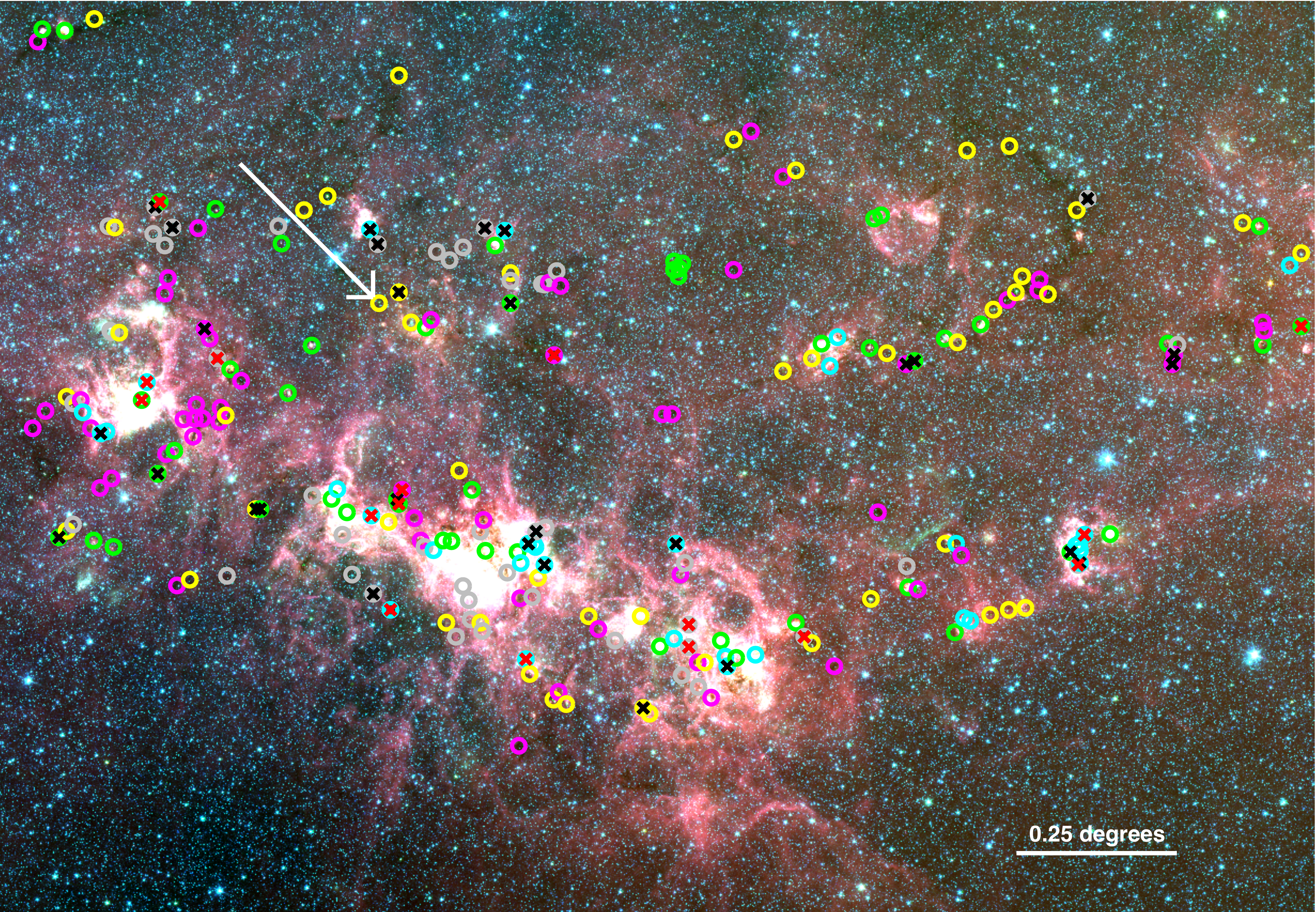}
\caption{$Spitzer$ IRAC 3-color image featuring the bright \ion{H}{2} complex G333. Red, green, and blue colors show 8, 4.5, and 3.6~$\mu$m respectively. The field is oriented in Galactic coordinates with a y-axis size of 1.4$^\circ$ and a center of $l=332.8^\circ$ and $b=-0.3^\circ$. MALT90 clump classifications from J. Rathborne et al. (in preparation) are shown, where magenta=Quiescent, yellow=Protostellar, green=\ion{H}{2}~region, cyan=PDR, and grey=Unknown. Locations of N$_2$H$^+$ poor sources are shown with $\times$ marks, where the black $\times$ marks show locations of N$_2$H$^+$ poor sources with 0.3~$>$~[I(N$_2$H$^+$)/I(HCO$^+$)]~$>$~0.2 \citep[based on the criteria in][]{Hoq2013} and red $\times$ marks show locations of the most extreme \nnhp~poor sources ([I(N$_2$H$^+$)/I(HCO$^+$)]~$<$~0.2) based on the criteria outlined in Section \ref{sec:MALT90n2hppoor}. The white arrow shows the location of G333.234--00.061, an \nthp\ rich source observed with the ATCA in this paper. The bright \ion{H}{2} complex is G333.  \label{glm333}
}
\end{center}
\end{figure*}

\begin{figure*} [ht!]Ê
\begin{center}
\includegraphics[scale=0.3]{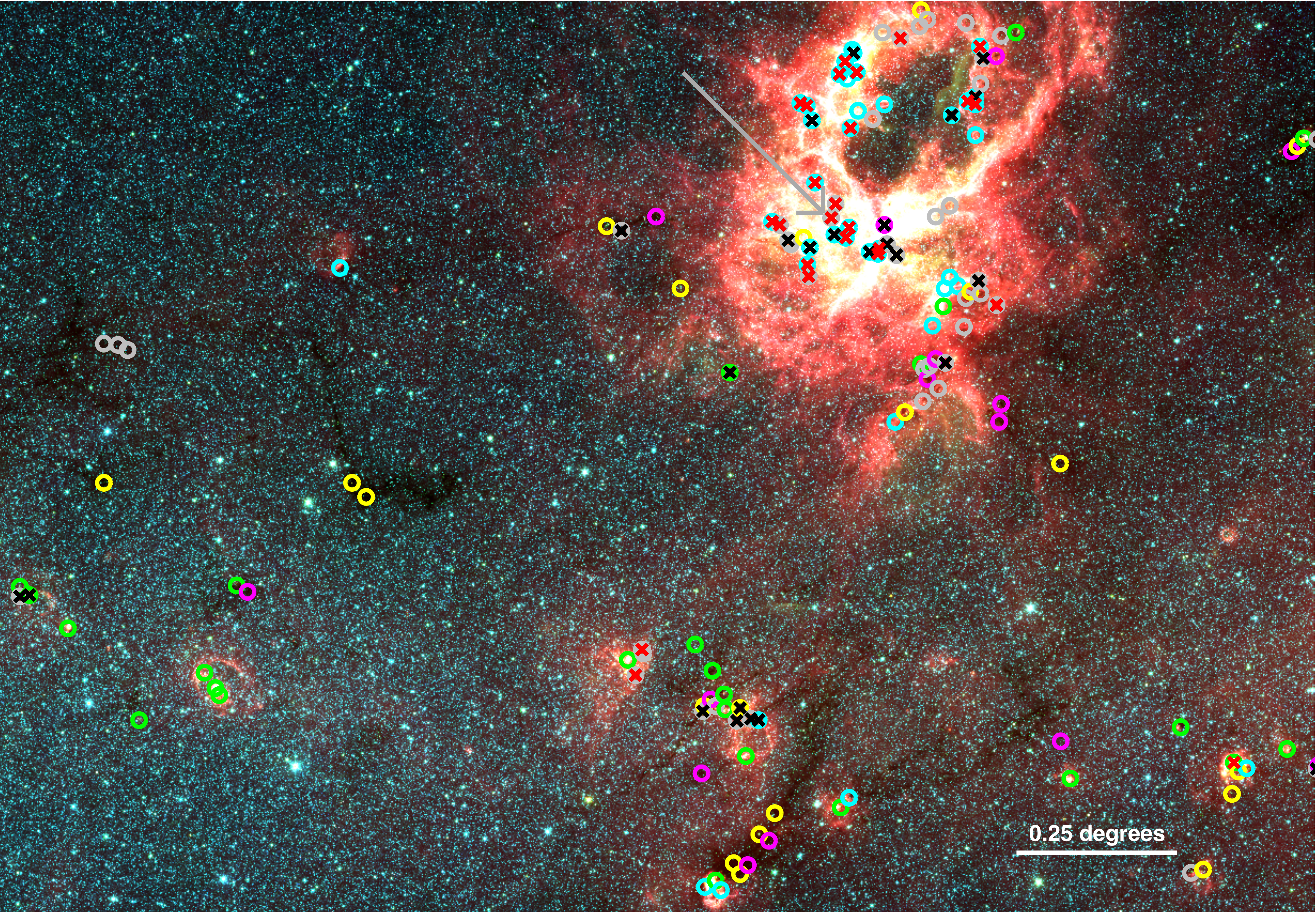}
\caption{$Spitzer$ IRAC 3-color image featuring the bright \ion{H}{2} complex NGC 6357. Caption is the same as Figure \ref{glm333} except with the field centered at $l=353.5^\circ$ and $b=0.3^\circ$. The white arrow shows the location of G353.229+00.672, an \nthp\ rich poor observed with the ATCA in this paper.  \label{glm353}
}
\end{center}
\end{figure*}

These 124 N$_2$H$^+$ poor clumps are especially concentrated in two regions: the G333 complex (Figure \ref{glm333}, showing 16 sources) and NGC 6357 (Figure \ref{glm353}, showing 26 sources including G353.229+00.672). Immediately it is evident from these Figures that all N$_2$H$^+$ poor sources are located on or very near \ion{H}{2} regions.

\begin{figure*} [ht!]Ê
\begin{center}
\includegraphics[scale=0.3]{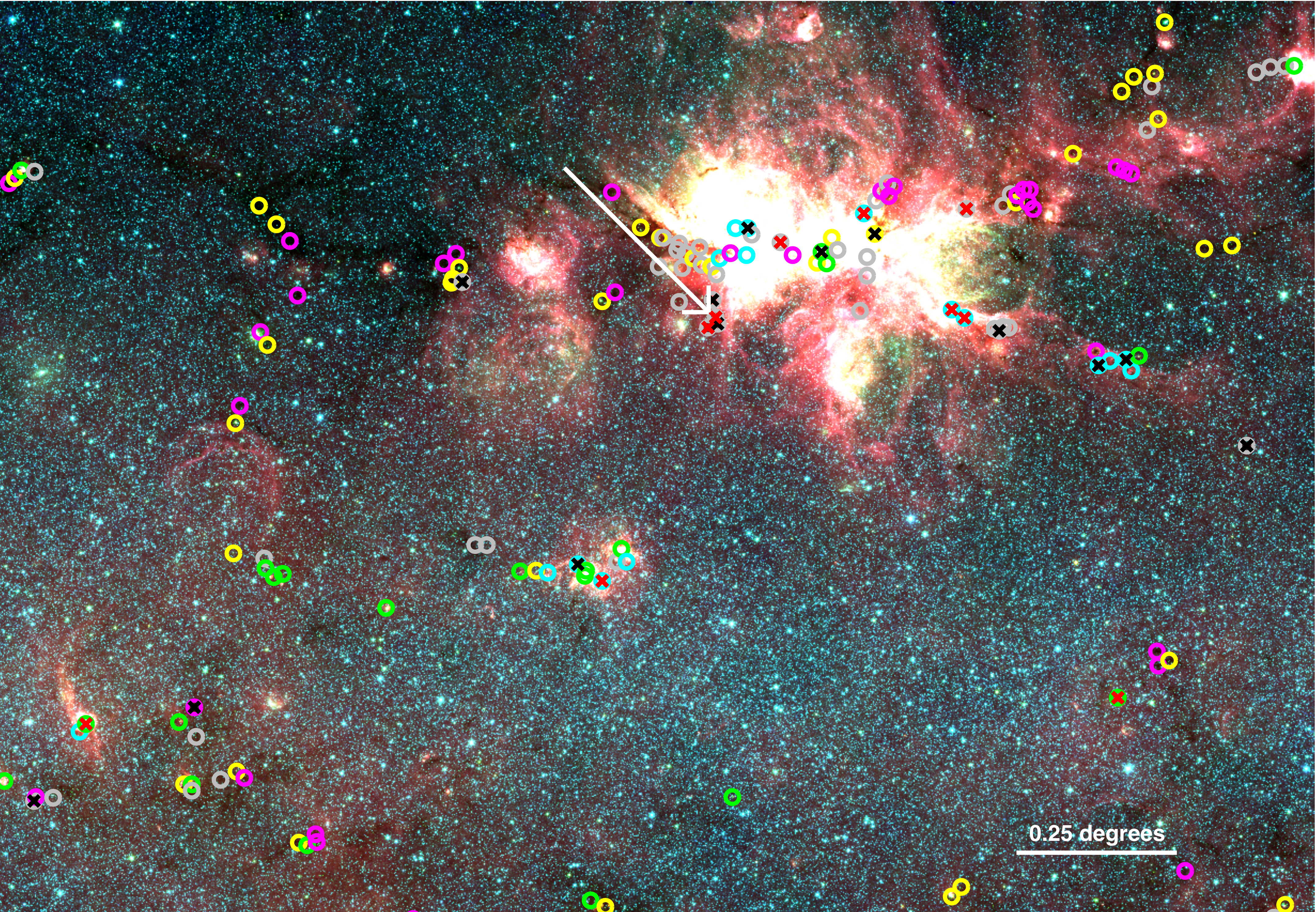}
\caption{$Spitzer$ IRAC 3-color image featuring the bright \ion{H}{2} complex NGC 6334. Caption is the same as Figure \ref{glm333} except with the field centered at $l=351.5^\circ$ and $b=0.35^\circ$. The gray arrow shows the location of G351.409+00.567, an \nthp\ poor source observed with the ATCA in this paper. \label{glm351}
}
\end{center}
\end{figure*}

Why does NGC~6357 contain so many (about one fifth) of these N$_2$H$^+$ poor clumps? One of the reasons is due to the fact that it is one of the closest ($\sim$1.9~kpc) high-mass star-forming regions in the Galaxy. This close proximity allows clumps to be resolved that would be blended at farther distances (Figure \ref{glm353} shows that many \nnhp~poor clumps are spatially near each other) and causes the clumps to be brighter due to the inverse square law. On the other hand, NGC~6334 (Figure \ref{glm351}, showing ten \nnhp~poor clumps) is also located at a close distance ($\sim$1.7~kpc), has a similar size, and is brighter than NGC~6357 but contains almost a factor of three fewer N$_2$H$^+$ poor clumps. The main difference between NGC~6334 and NGC~6357 is that NGC~6357 has a giant shell and the N$_2$H$^+$ poor clumps are coincident with this shell. The high-mass star-forming region G333 is also relatively close (a typical clump has a kinematic distance of 3.2~kpc, S. Whitaker et al., in preparation) and also has the majority of its N$_2$H$^+$ poor clumps bordering shells. These shells are typically the photodissociated interfaces between molecular and ionized gas. Thus, this analysis suggests N$_2$H$^+$ poor clumps are particularly found in PDR shells.


Although \nnhp~poor clumps are primarily associated with clumps classified as \ion{H}{2} regions and PDRs, there remain five \nnhp~poor clumps classified as protostellar and eight as quiescent. These 13 sources are all located in projection against bright mid-IR $Spitzer$ emission, and all but three sources have at least a marginal detection in C$_2$H (a PDR tracer). For the three sources undetected in C$_2$H, one source had a noisy spectrum and the other two were relatively weak sources (I(HCO$^+$)~$<$~3), and if observations were deeper, they may have had C$_2$H detections as well. The classification scheme is subject to errors due to the projection of multiple sources along the line of sight.  Moreover, it can often fail for distant sources due to foreground contamination. It is entirely possible that these 13 sources are in fact associated with PDRs or \ion{H}{2} regions within the 38$\arcsec$ Mopra beam.

Since \nnhp\ poor clumps in \ion{H}{2} and PDR regions are associated with large amounts of radiation, these clumps are likely heated to higher temperatures than the ``quiescent" or ``protostellar" clumps.  In higher temperature gas (warmer than the sublimation temperature of ices), CO is released from grains into the gas phase.  Since chemical reactions with CO both create HCO$^+$ and destroy N$_2$H$^+$ (as discussed in Section \ref{sec:intro}), warmer regions are expected to have a larger HCO$^+$/N$_2$H$^+$ abundance ratio. However, at least in some sources such as the two observed ATCA clumps G351.409+00.567 and G353.229+00.672, the anomaly is no longer apparent at the core scale. At this scale, the high-density material may be at a colder gas temperature due to shielding of incoming radiation allowing N$_2$H$^+$ to survive. 

\section{Summary} \label{sec:summary}
Based on MALT90 data, \citet{Hoq2013} identified N$_2$H$^+$ anomalous sources: sources with either high or low integrated intensity ratios between the N$_2$H$^+$(1--0) and HCO$^+$(1--0) lines. New ATCA observations of two N$_2$H$^+$ rich clumps and N$_2$H$^+$ poor clumps show that:

\begin{itemize}
\item In the two observed N$_2$H$^+$ rich sources, the N$_2$H$^+$ rich anomaly is due to extreme self-absorption of the HCO$^+$(1--0) line. Rather than an unusually high N$_2$H$^+$ to HCO$^+$ abundance ratio, the HCO$^+$ emission is suppressed because it is extremely optically thick.
\item The ATCA observations of the two observed N$_2$H$^+$ poor sources suggest this emission is anomalous on large (clump) scales rather than small (core) scales. The higher density of cores may help in shielding it from radiation, allowing \nnhp~to survive on small scales.
\item N$_2$H$^+$ poor clumps are primarily associated with PDRs and \ion{H}{2} regions, especially at locations of shell structures. At high temperatures, CO is more abundant, creating HCO$^+$ and destroying \nnhp.
\item Given the criteria in Section \ref{sec:MALT90n2hppoor}, NGC 6357 has more N$_2$H$^+$ poor star-forming clumps than any region surveyed by MALT90. It contains a plethora of these sources due to its close distance and the presence of a giant PDR shell.
\\

The \nnhp ``rich" clump G333.234--00.061 is an unusual star-forming region. Specifically:
\item It contains two of the most massive protostellar cores known separated by a projected distance of only 0.12~pc. These cores have estimated masses of at least 36~$M_\sun$ and 29~$M_\sun$ and diameters less than $\sim$0.06~pc.
\item Even though class I methanol masers are considered to indicate the location of outflows, the strongest outflow (ejected from MM2) is not associated with a class I methanol maser, while the core without a definite outflow, MM1, is associated with this maser. MM1's class I methanol maser might indicate an outflow forming in MM1 and generating strong collisions as it starts to break through its natal environment.
\item The more massive protostellar core, MM1, appears to be at an earlier evolutionary stage than MM2 (since MM2 has IRAC emission, a large and strong outflow, stronger $^{13}$CS emission, and later-stage masers). For two high-mass protostars that formed in the same molecular clump, it is unexpected to have the lower mass protostar evolving faster than the higher mass protostar. We suggest several interpretations that might explain this apparent discrepancy.


\end{itemize}

\acknowledgments
This work was supported by NASA grant NNX12AE42G and NSF grant AST-1211844. We thank Francesca Schiavello for assisting with the ATCA observations. The Australia Telescope Compact Array is part of the Australia Telescope National Facility which is funded by the Commonwealth of Australia for operation as a National Facility managed by CSIRO. This research made use of APLpy and pyspeckit, open-source plotting packages for Python hosted at http://aplpy.github.com and http://pyspeckit.bitbucket.org respectively. 

\newpage
\bibliography{stephens_bib}

\begin{thebibliography}{}
\expandafter\ifx\csname natexlab\endcsname\relax\def\natexlab#1{#1}\fi

\bibitem[{{Benjamin} {et~al.}(2003){Benjamin}, {Churchwell}, {Babler}, {Bania},
  {Clemens}, {Cohen}, {Dickey}, {Indebetouw}, {Jackson}, {Kobulnicky},
  {Lazarian}, {Marston}, {Mathis}, {Meade}, {Seager}, {Stolovy}, {Watson},
  {Whitney}, {Wolff}, \& {Wolfire}}]{Benjamin2003}
{Benjamin}, R.~A., {Churchwell}, E., {Babler}, B.~L., {et~al.} 2003, \pasp,
  115, 953

\bibitem[{{Bergin} \& {Langer}(1997)}]{Bergin1997}
{Bergin}, E.~A., \& {Langer}, W.~D. 1997, \apj, 486, 316

\bibitem[{{Breen} {et~al.}(2010){Breen}, {Caswell}, {Ellingsen}, \&
  {Phillips}}]{Breen2010}
{Breen}, S.~L., {Caswell}, J.~L., {Ellingsen}, S.~P., \& {Phillips}, C.~J.
  2010, \mnras, 406, 1487

\bibitem[{{Breen} {et~al.}(2007){Breen}, {Ellingsen}, {Johnston-Hollitt},
  {Wotherspoon}, {Bains}, {Burton}, {Cunningham}, {Lo}, {Senkbeil}, \&
  {Wong}}]{Breen2007}
{Breen}, S.~L., {Ellingsen}, S.~P., {Johnston-Hollitt}, M., {et~al.} 2007,
  \mnras, 377, 491

\bibitem[{{Brooks} {et~al.}(2007){Brooks}, {Garay}, {Voronkov}, \&
  {Rodr{\'{\i}}guez}}]{Brooks2007}
{Brooks}, K.~J., {Garay}, G., {Voronkov}, M., \& {Rodr{\'{\i}}guez}, L.~F.
  2007, \apj, 669, 459

\bibitem[{{Busquet} {et~al.}(2011){Busquet}, {Estalella}, {Zhang}, {Viti},
  {Palau}, {Ho}, \& {S{\'a}nchez-Monge}}]{Busquet2011}
{Busquet}, G., {Estalella}, R., {Zhang}, Q., {et~al.} 2011, \aap, 525, A141

\bibitem[{{Carey} {et~al.}(2009){Carey}, {Noriega-Crespo}, {Mizuno}, {Shenoy},
  {Paladini}, {Kraemer}, {Price}, {Flagey}, {Ryan}, {Ingalls}, {Kuchar},
  {Pinheiro Gon{\c c}alves}, {Indebetouw}, {Billot}, {Marleau}, {Padgett},
  {Rebull}, {Bressert}, {Ali}, {Molinari}, {Martin}, {Berriman}, {Boulanger},
  {Latter}, {Miville-Deschenes}, {Shipman}, \& {Testi}}]{Carey2009}
{Carey}, S.~J., {Noriega-Crespo}, A., {Mizuno}, D.~R., {et~al.} 2009, \pasp,
  121, 76

\bibitem[{{Caswell}(1998)}]{Caswell1998}
{Caswell}, J.~L. 1998, \mnras, 297, 215

\bibitem[{{Caswell} {et~al.}(2011){Caswell}, {Fuller}, {Green}, {Avison},
  {Breen}, {Ellingsen}, {Gray}, {Pestalozzi}, {Quinn}, {Thompson}, \&
  {Voronkov}}]{Caswell2011}
{Caswell}, J.~L., {Fuller}, G.~A., {Green}, J.~A., {et~al.} 2011, \mnras, 417,
  1964

\bibitem[{{Chambers} {et~al.}(2009){Chambers}, {Jackson}, {Rathborne}, \&
  {Simon}}]{Chambers2009}
{Chambers}, E.~T., {Jackson}, J.~M., {Rathborne}, J.~M., \& {Simon}, R. 2009,
  \apjs, 181, 360

\bibitem[{{Charnley}(1997)}]{Charnley1997}
{Charnley}, S.~B. 1997, \mnras, 291, 455

\bibitem[{{Cragg} {et~al.}(1992){Cragg}, {Johns}, {Godfrey}, \&
  {Brown}}]{Cragg1992}
{Cragg}, D.~M., {Johns}, K.~P., {Godfrey}, P.~D., \& {Brown}, R.~D. 1992,
  \mnras, 259, 203

\bibitem[{{Cragg} {et~al.}(2002){Cragg}, {Sobolev}, \& {Godfrey}}]{Cragg2002}
{Cragg}, D.~M., {Sobolev}, A.~M., \& {Godfrey}, P.~D. 2002, \mnras, 331, 521

\bibitem[{{Cyganowski} {et~al.}(2009){Cyganowski}, {Brogan}, {Hunter}, \&
  {Churchwell}}]{Cyganowski2009}
{Cyganowski}, C.~J., {Brogan}, C.~L., {Hunter}, T.~R., \& {Churchwell}, E.
  2009, \apj, 702, 1615

\bibitem[{{Cyganowski} {et~al.}(2011){Cyganowski}, {Brogan}, {Hunter}, \&
  {Churchwell}}]{Cyganowski2011}
---. 2011, \apj, 743, 56

\bibitem[{{Cyganowski} {et~al.}(2012){Cyganowski}, {Brogan}, {Hunter}, {Zhang},
  {Friesen}, {Indebetouw}, \& {Chandler}}]{Cyganowski2012}
{Cyganowski}, C.~J., {Brogan}, C.~L., {Hunter}, T.~R., {et~al.} 2012, \apjl,
  760, L20

\bibitem[{{Daniel} {et~al.}(2006){Daniel}, {Cernicharo}, \&
  {Dubernet}}]{Daniel2006}
{Daniel}, F., {Cernicharo}, J., \& {Dubernet}, M.-L. 2006, \apj, 648, 461

\bibitem[{{Elitzur} {et~al.}(1989){Elitzur}, {Hollenbach}, \&
  {McKee}}]{Elitzur1989}
{Elitzur}, M., {Hollenbach}, D.~J., \& {McKee}, C.~F. 1989, \apj, 346, 983

\bibitem[{{Ellingsen}(2005)}]{Ellingsen2005}
{Ellingsen}, S.~P. 2005, \mnras, 359, 1498

\bibitem[{{Ellingsen}(2006)}]{Ellingsen2006}
---. 2006, \apj, 638, 241

\bibitem[{{Fern{\'a}ndez-L{\'o}pez} {et~al.}(2011){Fern{\'a}ndez-L{\'o}pez},
  {Curiel}, {Girart}, {Ho}, {Patel}, \& {G{\'o}mez}}]{FL2011}
{Fern{\'a}ndez-L{\'o}pez}, M., {Curiel}, S., {Girart}, J.~M., {et~al.} 2011,
  \aj, 141, 72

\bibitem[{{Forster} \& {Caswell}(1989)}]{Forster1989}
{Forster}, J.~R., \& {Caswell}, J.~L. 1989, \aap, 213, 339

\bibitem[{{Foster} {et~al.}(2011){Foster}, {Jackson}, {Barnes}, {Barris},
  {Brooks}, {Cunningham}, {Finn}, {Fuller}, {Longmore}, {Mascoop}, {Peretto},
  {Rathborne}, {Sanhueza}, {Schuller}, \& {Wyrowski}}]{Foster2011}
{Foster}, J.~B., {Jackson}, J.~M., {Barnes}, P.~J., {et~al.} 2011, \apjs, 197,
  25

\bibitem[{{Foster} {et~al.}(2013){Foster}, {Rathborne}, {Sanhueza},
  {Claysmith}, {Whitaker}, {Jackson}, {Mascoop}, {Wienen}, {Breen}, {Herpin},
  {Duarte-Cabral}, {Csengeri}, {Contreras}, {Indermuehle}, {Barnes}, {Walsh},
  {Cunningham}, {Britton}, {Voronkov}, {Urquhart}, {Alves}, {Jordan}, {Hill},
  {Hoq}, {Brooks}, \& {Longmore}}]{Foster2013}
{Foster}, J.~B., {Rathborne}, J.~M., {Sanhueza}, P., {et~al.} 2013, \pasa, 30,
  38

\bibitem[{{Franco} {et~al.}(2000){Franco}, {Kurtz}, {Hofner}, {Testi},
  {Garc{\'{\i}}a-Segura}, \& {Martos}}]{Franco2000}
{Franco}, J., {Kurtz}, S., {Hofner}, P., {et~al.} 2000, \apjl, 542, L143

\bibitem[{{Gan} {et~al.}(2013){Gan}, {Chen}, {Shen}, {Xu}, \& {Ju}}]{Gan2013}
{Gan}, C.-G., {Chen}, X., {Shen}, Z.-Q., {Xu}, Y., \& {Ju}, B.-G. 2013, \apj,
  763, 2

\bibitem[{{Hildebrand}(1983)}]{Hildebrand1983}
{Hildebrand}, R.~H. 1983, \qjras, 24, 267

\bibitem[{{Hoare} {et~al.}(2007){Hoare}, {Kurtz}, {Lizano}, {Keto}, \&
  {Hofner}}]{Hoare2007}
{Hoare}, M.~G., {Kurtz}, S.~E., {Lizano}, S., {Keto}, E., \& {Hofner}, P. 2007,
  Protostars and Planets V, 181

\bibitem[{{Hoq} {et~al.}(2013){Hoq}, {Jackson}, {Foster}, {Sanhueza},
  {Guzm{\'a}n}, {Whitaker}, {Claysmith}, {Rathborne}, {Vasyunina}, \&
  {Vasyunin}}]{Hoq2013}
{Hoq}, S., {Jackson}, J.~M., {Foster}, J.~B., {et~al.} 2013, \apj, 777, 157

\bibitem[{{Jackson} {et~al.}(2013){Jackson}, {Rathborne}, {Foster}, {Whitaker},
  {Sanhueza}, {Claysmith}, {Mascoop}, {Wienen}, {Breen}, {Herpin},
  {Duarte-Cabral}, {Csengeri}, {Longmore}, {Contreras}, {Indermuehle},
  {Barnes}, {Walsh}, {Cunningham}, {Brooks}, {Britton}, {Voronkov}, {Urquhart},
  {Alves}, {Jordan}, {Hill}, {Hoq}, {Finn}, {Bains}, {Bontemps}, {Bronfman},
  {Caswell}, {Deharveng}, {Ellingsen}, {Fuller}, {Garay}, {Green}, {Hindson},
  {Jones}, {Lenfestey}, {Lo}, {Lowe}, {Mardones}, {Menten}, {Minier}, {Morgan},
  {Motte}, {Muller}, {Peretto}, {Purcell}, {Schilke}, {Bontemps}, {Schuller},
  {Titmarsh}, {Wyrowski}, \& {Zavagno}}]{Jackson2013}
{Jackson}, J.~M., {Rathborne}, J.~M., {Foster}, J.~B., {et~al.} 2013, \pasa,
  30, 57

\bibitem[{{Jansen} {et~al.}(1995){Jansen}, {van Dishoeck}, {Black}, {Spaans},
  \& {Sosin}}]{Jansen1995}
{Jansen}, D.~J., {van Dishoeck}, E.~F., {Black}, J.~H., {Spaans}, M., \&
  {Sosin}, C. 1995, \aap, 302, 223

\bibitem[{{Kauffmann} {et~al.}(2008){Kauffmann}, {Bertoldi}, {Bourke}, {Evans},
  \& {Lee}}]{Kauffmann2008}
{Kauffmann}, J., {Bertoldi}, F., {Bourke}, T.~L., {Evans}, II, N.~J., \& {Lee},
  C.~W. 2008, \aap, 487, 993

\bibitem[{{Kurtz} {et~al.}(2004){Kurtz}, {Hofner}, \&
  {{\'A}lvarez}}]{Kurtz2004}
{Kurtz}, S., {Hofner}, P., \& {{\'A}lvarez}, C.~V. 2004, \apjs, 155, 149

\bibitem[{{Leurini} {et~al.}(2004){Leurini}, {Schilke}, {Menten}, {Flower},
  {Pottage}, \& {Xu}}]{Leurini2004}
{Leurini}, S., {Schilke}, P., {Menten}, K.~M., {et~al.} 2004, \aap, 422, 573

\bibitem[{{Molinari} {et~al.}(2010){Molinari}, {Swinyard}, {Bally}, {Barlow},
  {Bernard}, {Martin}, {Moore}, {Noriega-Crespo}, {Plume}, {Testi}, {Zavagno},
  {Abergel}, {Ali}, {Anderson}, {Andr{\'e}}, {Baluteau}, {Battersby},
  {Beltr{\'a}n}, {Benedettini}, {Billot}, {Blommaert}, {Bontemps}, {Boulanger},
  {Brand}, {Brunt}, {Burton}, {Calzoletti}, {Carey}, {Caselli}, {Cesaroni},
  {Cernicharo}, {Chakrabarti}, {Chrysostomou}, {Cohen}, {Compiegne}, {de
  Bernardis}, {de Gasperis}, {di Giorgio}, {Elia}, {Faustini}, {Flagey},
  {Fukui}, {Fuller}, {Ganga}, {Garcia-Lario}, {Glenn}, {Goldsmith}, {Griffin},
  {Hoare}, {Huang}, {Ikhenaode}, {Joblin}, {Joncas}, {Juvela}, {Kirk},
  {Lagache}, {Li}, {Lim}, {Lord}, {Marengo}, {Marshall}, {Masi}, {Massi},
  {Matsuura}, {Minier}, {Miville-Desch{\^e}nes}, {Montier}, {Morgan}, {Motte},
  {Mottram}, {M{\"u}ller}, {Natoli}, {Neves}, {Olmi}, {Paladini}, {Paradis},
  {Parsons}, {Peretto}, {Pestalozzi}, {Pezzuto}, {Piacentini}, {Piazzo},
  {Polychroni}, {Pomar{\`e}s}, {Popescu}, {Reach}, {Ristorcelli}, {Robitaille},
  {Robitaille}, {Rod{\'o}n}, {Roy}, {Royer}, {Russeil}, {Saraceno}, {Sauvage},
  {Schilke}, {Schisano}, {Schneider}, {Schuller}, {Schulz}, {Sibthorpe},
  {Smith}, {Smith}, {Spinoglio}, {Stamatellos}, {Strafella}, {Stringfellow},
  {Sturm}, {Taylor}, {Thompson}, {Traficante}, {Tuffs}, {Umana}, {Valenziano},
  {Vavrek}, {Veneziani}, {Viti}, {Waelkens}, {Ward-Thompson}, {White},
  {Wilcock}, {Wyrowski}, {Yorke}, \& {Zhang}}]{Molinari2010}
{Molinari}, S., {Swinyard}, B., {Bally}, J., {et~al.} 2010, \aap, 518, L100

\bibitem[{{Osorio} {et~al.}(1999){Osorio}, {Lizano}, \&
  {D'Alessio}}]{Osorio1999}
{Osorio}, M., {Lizano}, S., \& {D'Alessio}, P. 1999, \apj, 525, 808

\bibitem[{{Ossenkopf} \& {Henning}(1994)}]{Ossenkopf1994}
{Ossenkopf}, V., \& {Henning}, T. 1994, \aap, 291, 943

\bibitem[{{Peretto} {et~al.}(2013){Peretto}, {Fuller}, {Duarte-Cabral},
  {Avison}, {Hennebelle}, {Pineda}, {Andr{\'e}}, {Bontemps}, {Motte},
  {Schneider}, \& {Molinari}}]{Peretto2013}
{Peretto}, N., {Fuller}, G.~A., {Duarte-Cabral}, A., {et~al.} 2013, \aap, 555,
  A112

\bibitem[{{Pihlstr{\"o}m} {et~al.}(2014){Pihlstr{\"o}m}, {Sjouwerman}, {Frail},
  {Claussen}, {Mesler}, \& {McEwen}}]{Pihlstrom2014}
{Pihlstr{\"o}m}, Y.~M., {Sjouwerman}, L.~O., {Frail}, D.~A., {et~al.} 2014,
  \aj, 147, 73

\bibitem[{{Quireza} {et~al.}(2006){Quireza}, {Rood}, {Bania}, {Balser}, \&
  {Maciel}}]{Quireza2006}
{Quireza}, C., {Rood}, R.~T., {Bania}, T.~M., {Balser}, D.~S., \& {Maciel},
  W.~J. 2006, \apj, 653, 1226

\bibitem[{{Remijan} {et~al.}(2007){Remijan}, {Markwick-Kemper}, \& {ALMA
  Working Group on Spectral Line Frequencies}}]{Remijan2007}
{Remijan}, A.~J., {Markwick-Kemper}, A., \& {ALMA Working Group on Spectral
  Line Frequencies}. 2007, in Bulletin of the American Astronomical Society,
  Vol.~39, American Astronomical Society Meeting Abstracts, 132.11

\bibitem[{{Russeil} {et~al.}(2012){Russeil}, {Zavagno}, {Adami}, {Anderson},
  {Bontemps}, {Motte}, {Rodon}, {Schneider}, {Ilmane}, \&
  {Murphy}}]{Russeil2012}
{Russeil}, D., {Zavagno}, A., {Adami}, C., {et~al.} 2012, \aap, 538, A142

\bibitem[{{Sanhueza} {et~al.}(2012){Sanhueza}, {Jackson}, {Foster}, {Garay},
  {Silva}, \& {Finn}}]{Sanhueza2012}
{Sanhueza}, P., {Jackson}, J.~M., {Foster}, J.~B., {et~al.} 2012, \apj, 756, 60

\bibitem[{{Sanhueza} {et~al.}(2013){Sanhueza}, {Jackson}, {Foster},
  {Jimenez-Serra}, {Dirienzo}, \& {Pillai}}]{Sanhueza2013}
---. 2013, \apj, 773, 123

\bibitem[{{Sault} {et~al.}(1995){Sault}, {Teuben}, \& {Wright}}]{Sault1995}
{Sault}, R.~J., {Teuben}, P.~J., \& {Wright}, M.~C.~H. 1995, in Astronomical
  Society of the Pacific Conference Series, Vol.~77, Astronomical Data Analysis
  Software and Systems IV, ed. R.~A. {Shaw}, H.~E. {Payne}, \& J.~J.~E.
  {Hayes}, 433

\bibitem[{{Schuller} {et~al.}(2009){Schuller}, {Menten}, {Contreras},
  {Wyrowski}, {Schilke}, {Bronfman}, {Henning}, {Walmsley}, {Beuther},
  {Bontemps}, {Cesaroni}, {Deharveng}, {Garay}, {Herpin}, {Lefloch}, {Linz},
  {Mardones}, {Minier}, {Molinari}, {Motte}, {Nyman}, {Reveret}, {Risacher},
  {Russeil}, {Schneider}, {Testi}, {Troost}, {Vasyunina}, {Wienen}, {Zavagno},
  {Kovacs}, {Kreysa}, {Siringo}, \& {Wei{\ss}}}]{Schuller2009}
{Schuller}, F., {Menten}, K.~M., {Contreras}, Y., {et~al.} 2009, \aap, 504, 415

\bibitem[{{Tan} {et~al.}(2014){Tan}, {Beltran}, {Caselli}, {Fontani}, {Fuente},
  {Krumholz}, {McKee}, \& {Stolte}}]{Tan2014}
{Tan}, J.~C., {Beltran}, M.~T., {Caselli}, P., {et~al.} 2014, ArXiv e-prints,
  arXiv:1402.0919

\bibitem[{{Voronkov} {et~al.}(2006){Voronkov}, {Brooks}, {Sobolev},
  {Ellingsen}, {Ostrovskii}, \& {Caswell}}]{Voronkov2006}
{Voronkov}, M.~A., {Brooks}, K.~J., {Sobolev}, A.~M., {et~al.} 2006, \mnras,
  373, 411

\bibitem[{{Voronkov} {et~al.}(2014){Voronkov}, {Caswell}, {Ellingsen}, {Green},
  \& {Breen}}]{Voronkov2014}
{Voronkov}, M.~A., {Caswell}, J.~L., {Ellingsen}, S.~P., {Green}, J.~A., \&
  {Breen}, S.~L. 2014, \mnras, 439, 2584

\bibitem[{{Voronkov} {et~al.}(2010){Voronkov}, {Caswell}, {Ellingsen}, \&
  {Sobolev}}]{Voronkov2010}
{Voronkov}, M.~A., {Caswell}, J.~L., {Ellingsen}, S.~P., \& {Sobolev}, A.~M.
  2010, \mnras, 405, 2471

\bibitem[{{Xu} {et~al.}(2008){Xu}, {Li}, {Hachisuka}, {Pandian}, {Menten}, \&
  {Henkel}}]{Xu2008}
{Xu}, Y., {Li}, J.~J., {Hachisuka}, K., {et~al.} 2008, \aap, 485, 729

\bibitem[{{Zhang} {et~al.}(2014){Zhang}, {Wang}, {Xu}, {Wyrowski}, \&
  {Menten}}]{Zhang2014}
{Zhang}, C.-P., {Wang}, J.-J., {Xu}, J.-L., {Wyrowski}, F., \& {Menten}, K.~M.
  2014, \apj, 784, 107

\bibitem[{{Zinnecker} \& {Yorke}(2007)}]{Zinnecker2007}
{Zinnecker}, H., \& {Yorke}, H.~W. 2007, \araa, 45, 481

\end{thebibliography}
\end{document}